\documentclass[12pt]{iopart}
\pdfoutput=1
\expandafter\let\csname equation*\endcsname\relax
\expandafter\let\csname endequation*\endcsname\relax
\usepackage[numbers,sort&compress]{natbib}
\usepackage{amsmath,amssymb,eucal,graphicx,bm}
\usepackage{subfigure}
\usepackage{float}

\def\ltwid{\mathrel{\raise.3ex\hbox{$<$\kern-.75em\lower1ex\hbox{$\sim$}}}}

\usepackage{appendix}
\usepackage[breakwords]{truncate}
\usepackage{lastpage}
\usepackage{hyperref}
\usepackage{amsfonts}
\usepackage{physics}
\usepackage[section]{placeins}
\usepackage{color,url}
\def \Pe {\text{Pe}}
\def \P {\text{P}}
\def \Fo {\text{Fo}}

\def \I {\text{I}}
\def \K {\text{K}}

\begin{document}

\title{Optimization and Growth in First-Passage Resetting}
\author{B. De Bruyne}
\address{Perimeter Institute, 31 Caroline Street North, Waterloo, ON, N2L
  2Y5, Canada}
\address{LPTMS, CNRS, Univ.\ Paris-Sud, Universit\'e Paris-Saclay, 91405 Orsay, France}
\author{J. Randon-Furling}
\address{SAMM, Universit\'e Paris 1 -- FP2M (FR2036) CNRS, 75013 Paris, France}
\address{Department of Mathematics, Columbia University, New York, NY 10027, USA}
\address{MSDA, Mohammed VI Polytechnic University, Ben Guerir 43150, Morocco}
\author{S. Redner}
\address{Santa Fe Institute, 1399 Hyde Park Rd., Santa Fe, New Mexico 87501, USA}

\begin{abstract}
  We combine the processes of resetting and first-passage to define
  \emph{first-passage resetting}, where the resetting of a random walk to a
  fixed position is triggered by a first-passage event of the walk itself.
  In an infinite domain, first-passage resetting of isotropic diffusion is
  non-stationary, with the number of resetting events growing with time as
  $\sqrt{t}$.  We calculate the resulting spatial probability distribution of
  the particle analytically, and also obtain this distribution by a geometric
  path decomposition.  In a finite interval, we define an optimization
  problem that is controlled by first-passage resetting; this scenario is
  motivated by reliability theory.  The goal is to operate a system close to
  its maximum capacity without experiencing too many breakdowns.  However,
  when a breakdown occurs the system is reset to its minimal operating point.
  We define and optimize an objective function that maximizes the reward
  (being close to maximum operation) minus a penalty for each breakdown.  We
  also investigate extensions of this basic model to include delay after each
  reset and to two dimensions.  Finally, we study the growth dynamics of a
  domain in which the domain boundary recedes by a specified amount whenever
  the diffusing particle reaches the boundary after which a resetting event
  occurs.  We determine the growth rate of the domain for the semi-infinite
  line and the finite interval and find a wide range of behaviors that depend
  on how much the recession occurs when the particle hits the boundary.
\end{abstract}

\section{Introduction}

Random walks are ubiquitous in phenomena across a wide range of fields, such
as physics, chemistry, finance and social
sciences~\cite{chandrasekhar1943stochastic,van1992stochastic,chicheportiche2014some,rogers2010diffusion}.
In addition to the applications of the random walk, many useful extensions of
the basic model have been developed (see, e.g.,
\cite{weiss1994aspects,hughes1995random,rudnick2004elements,klafter2011first}).
Almost a decade ago, the notion of resetting of a random walk was
introduced~\cite{evans2011diffusion,evans2011diffusionjpa,evans2020stochastic}.
The basic idea of resetting is simplicity itself: at a given rate, a random
walk is reset to its starting point.  The rich phenomenology induced by this
extension of the random walk has sparked much interest (see,
e.g.,~\cite{evans2011diffusion,evans2011diffusionjpa,evans2020stochastic,PhysRevLett.112.240601,christou2015diffusion,PhysRevE.92.060101,PhysRevE.92.052126,reuveni2016optimal,pal2017first,belan2018restart,bodrova2019nonrenewal}).
In the context of search strategies, where the walker is searching for a
target at a fixed location, resetting changes the average search time from
being infinite (in an infinite domain) to
finite~\cite{PhysRevE.92.052127,PhysRevE.92.062115,bhat2016stochastic,eule2016non}.
Moreover, there exists an optimal resetting rate that minimizes the search
time.  A very different, but also fruitful concept in the theory of random
walks is the notion of a first-passage
process~\cite{feller2008introduction,van1992stochastic,redner2001guide,Bray13}.  Of
particular importance is first-passage probability, which is defined as the
probability that a walker reaches a specified location for the first
time. This notion has many applications where a particular event happens when
a threshold is first reached.  One such example is a limit order for a stock.
When the price of a stock, whose evolution is often modeled as a geometric
random walk, first reaches a limit price, this event triggers the sale or the
purchase of the stock.

In this work, we combine these disparate notions of first passage and
resetting into \emph{first-passage resetting}, in which the particle is reset
whenever it reaches a specified threshold.  Contrary to standard resetting,
the time at which first-passage resetting occurs is defined by the motion of
the diffusing particle itself rather than being imposed
externally~\cite{evans2011diffusion,evans2011diffusionjpa,evans2020stochastic}. Feller
showed that such a process is well defined mathematically and provided
existence and uniqueness theorems~\cite{feller1954diffusion}, while similar
ideas were pursued in~\cite{sherman1958limiting}.  First-passage resetting
was initially treated in the physics literature for the situation in which
two Brownian particles are biased toward each other and the particles are
reset to their initial positions when they encounter each
other~\cite{falcao2017interacting}.

In our work, we first focus on the related situation of a diffusing particle
on the semi-infinite line $x\leq L$ that is reset to the origin whenever the
particle hits the boundary $x=L$ (Fig.~\ref{fig:model}).  The model studied
in~\cite{falcao2017interacting} corresponds to a drift toward the boundary in
our semi-infinite geometry; this setting leads to a stationary state.  In
contrast, the absence of drift in our model leads to a variety of new
phenomena.  In particular, the probability distribution for the position of
the particle is non-stationary.  We also construct two simple path
decompositions for first-passage resetting, in which the trajectory of the
resetting particle is mapped onto a free diffusion process.  This approach
provides useful geometrical insights, as well as simple ways to derive the
average number of reset events and the spatial probability distribution with
essentially no calculation.

We then treat first-passage resetting on a finite interval, which has a
natural application to reliability theory.  Here the particle is restricted
to the interval $[0,L]$ where $x=L$ is again the boundary where resetting
occurs and the particle is immediately reinjected at $x=0$ when it reaches
$x=L$.  We may view this mechanism as characterizing the performance of a
driven mechanical
system~\cite{ross2014introduction,gnedenko2014mathematical,317656,867175},
with the coordinates $x=0$ and $x=L$ indicating poor and maximal performance,
respectively.  While one ideally wants to operate the system close to its
maximum performance level ($x=L$), there is a risk of overuse, leading to
breakdowns whenever $x=L$ is reached.  Subsequently, the system has to be
repaired and then restarted from $x=0$.  This dynamics corresponds to
resetting that is induced by a first passage to the boundary $x=L$.  We will
find the optimal bias velocity that optimizes the performance of the system.
We will also investigate additional features of this first-passage resetting,
such as a random maintenance delay at each breakdown and resetting in higher
dimensions.  A preliminary account of some of these results was given
in~\cite{bruyne2020firstpassage}.

Finally, we investigate a very different aspect of first-passage resetting
where the domain boundary at which resetting occurs moves by a specified
amount each time the diffusing particle reaches this boundary.  Many features
of this moving boundary problem can be readily calculated because of the
renewal structure of the theory.  For both the semi-infinite and finite
interval geometries, we find a variety of scaling behaviors for the motion of
the resetting boundary.  These behaviors depend on the initial geometry and
by how much the boundary moves at each resetting event.

\section{First-Passage Resetting in the Semi-Infinite Geometry}
\label{sec:semiInfiniteGeometry}

\begin{figure}[ht]
  \center{ \includegraphics[width=0.6\textwidth]{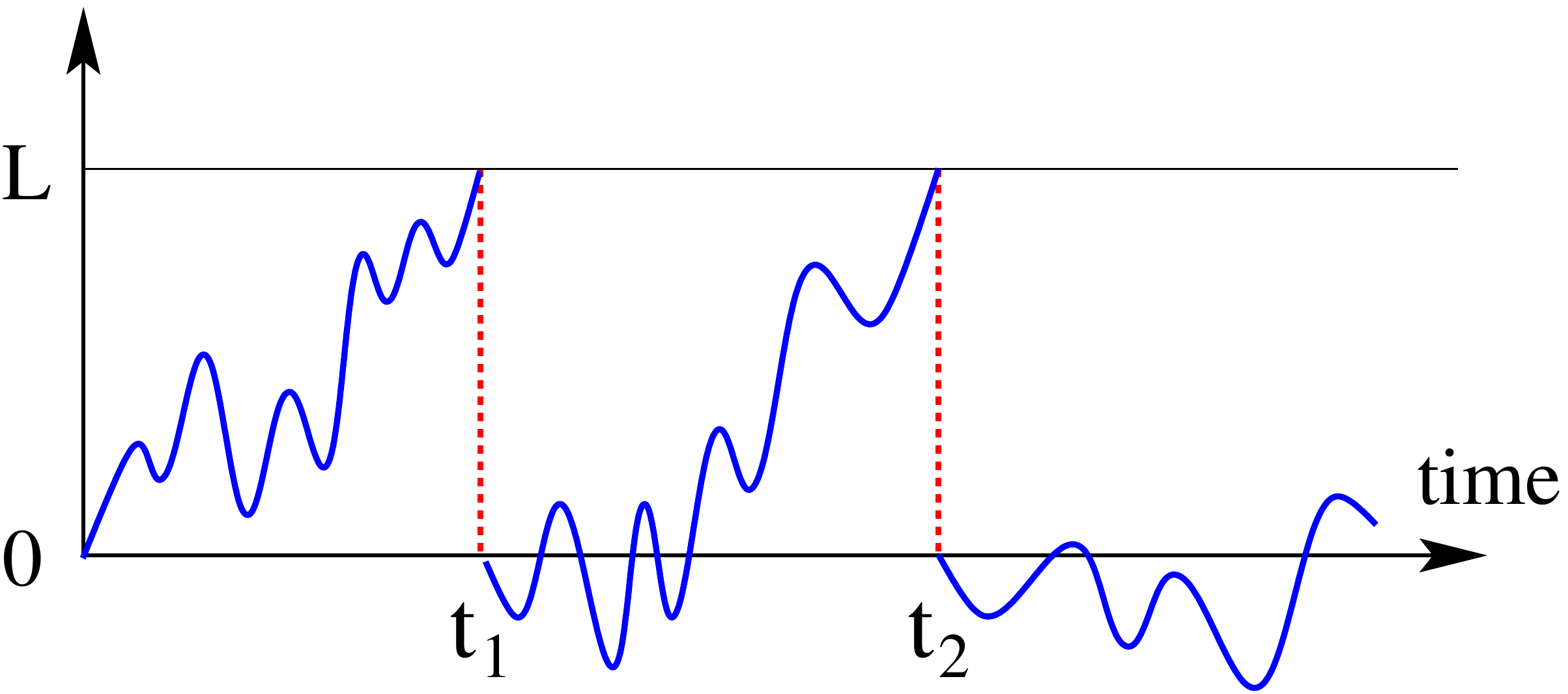}}
  \caption{Schematic illustration of first-passage resetting for diffusion on
    the semi-infinite line $x\leq L$.  Each time the particle reaches the
    threshold $L$, it is reset to the origin. The times of the resetting
    events are denoted by $t_1, t_2,\ldots$.}
\label{fig:model}
\end{figure}

In the standard resetting process, reset events occur at a fixed rate $r$ that
are uncorrelated with the position of the diffusing particle. In contrast,
first-passage resetting directly couples the times at which resetting occurs
and the particle position.  For first-passage resetting in the semi-infinite
line geometry, the particle starts at $x(0)=0$ and freely diffuses in the
range $x\leq L$ (with $L>0$).  Each time $L$ is reached, the particle is
instantaneously reset to the origin (Fig.~\ref{fig:model}).  We are
interested in two basic characteristics of the particle motion: the spatial
probability distribution of the particle and the time dependence of the
number of reset events.  To compute these quantities, we rely on the renewal
structure of the process, which allows us to first compute the probability
for $n$ reset events in a direct way.  With this result, as well as the
propagator for free diffusion in the presence of an absorbing boundary, we
can readily obtain the spatial distribution of the particle and the average
number of resets up to a given time.

\subsection{The $n^{\text th}$ reset probability distribution}

Define $F_n(L,t)$ as the probability that the particle resets for the
$n^{\text th}$ time at time $t$.  When $n=1$, this quantity is the standard
first-passage probability for a freely diffusing particle that starts at the
origin, to first reach $L$~\cite{redner2001guide, Bray13}:
\begin{align*}
F_1(L,t) \equiv F(L,t) = \frac{L}{\sqrt{4\pi D t^3}}\, e^{-L^2/4Dt}\,.
\end{align*}
For the particle to reset for the $n^{\text{th}}$ time at time $t$ with
$n>1$, the particle must reset for the $(n-1)^{\text{th}}$ time at some
earlier time $t'<t$, and reset one more time at time $t$.  Because the
process is renewed at each reset, $F_n(L,t)$ is formally given by the renewal
equation
\begin{subequations}
\begin{align}
\label{eq:Fnsi}
  F_n(L,t) = \int_0^t dt'\, F_{n-1}(L,t')\, F_1(L,t-t'),\qquad n>1\,.
\end{align}
The convolution structure of Eq.~\eqref{eq:Fnsi} lends itself to a Laplace
transform analysis. The corresponding equation in the Laplace domain
is simply:
\begin{align}
\label{eq:convFns}
    \widetilde F_n(L,s)=\widetilde F_{n-1}(L,s) \widetilde
  F_1(L,s)= \widetilde F_1(L,s)^n,
\end{align}
\end{subequations}
where quantities with tildes denote Laplace transforms.  Using the Laplace
transform of the first-passage probability:
\begin{align*}
  \widetilde F_1(L,s)=\int_0^\infty dt\, F_1(L,t)\, e^{-s\, t} = e^{\sqrt{sL^2/D}}\equiv
  e^{-\ell}\, ,
\end{align*}
where we define the scaled coordinate $\ell\equiv \sqrt{s/D}\,L$ henceforth,
then Eq.~\eqref{eq:convFns} becomes
\begin{align*}
  \widetilde F_n(L,s)= e^{-n\ell}\, .
\end{align*}
Notice that $\widetilde F_n(L,s)$ has the same form as $\widetilde F_1(L,s)$
with $L\to nL$.  That is, the time for a diffusing particle to reset $n$
times at a fixed boundary $x=L$ is the same as the time for a freely
diffusing particle to first reach $x=nL$.

\begin{figure}[ht]
  \centerline{ \vbox{\includegraphics[width=0.475\textwidth]{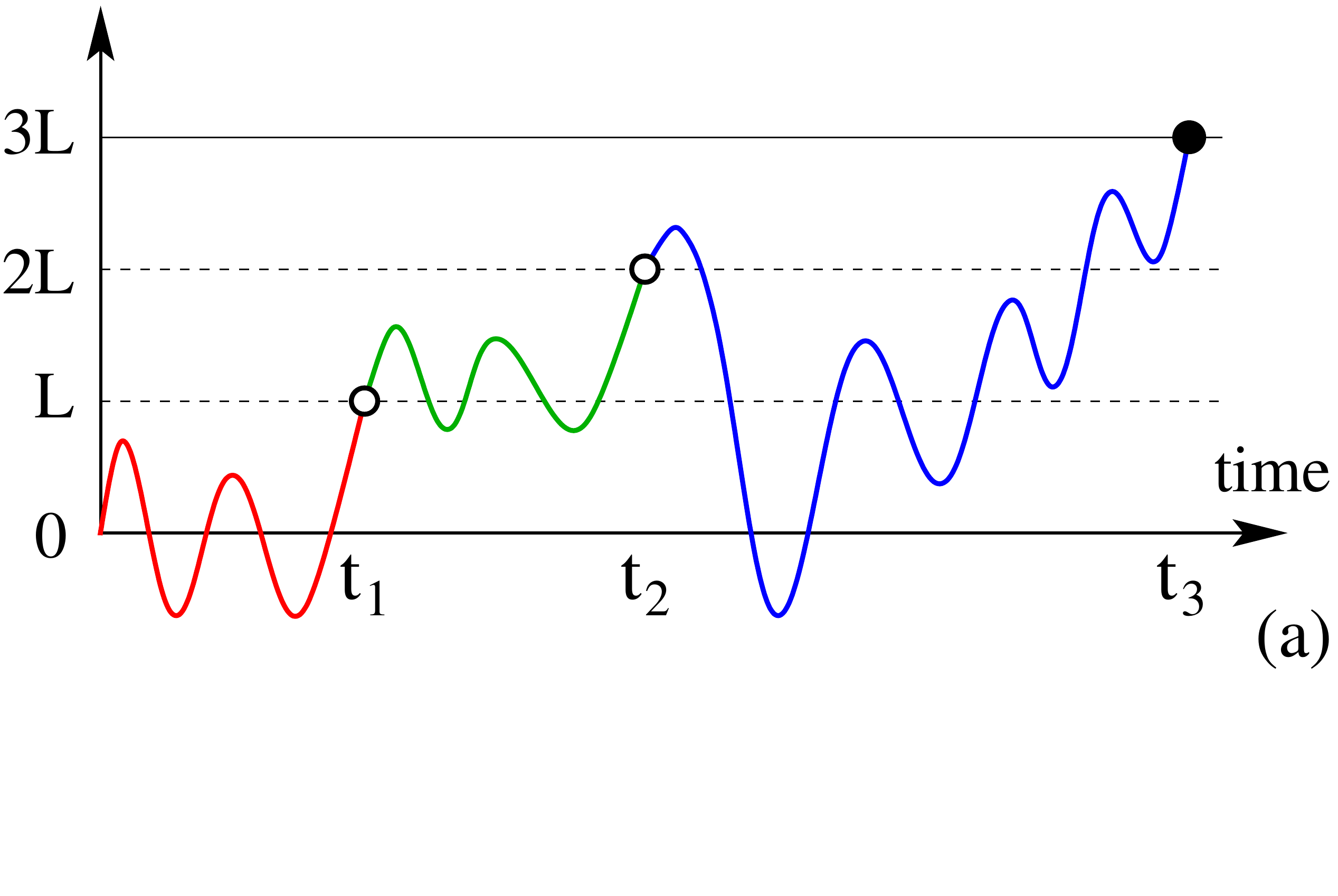}\qquad
  \vspace*{-0cm}\includegraphics[width=0.475\textwidth]{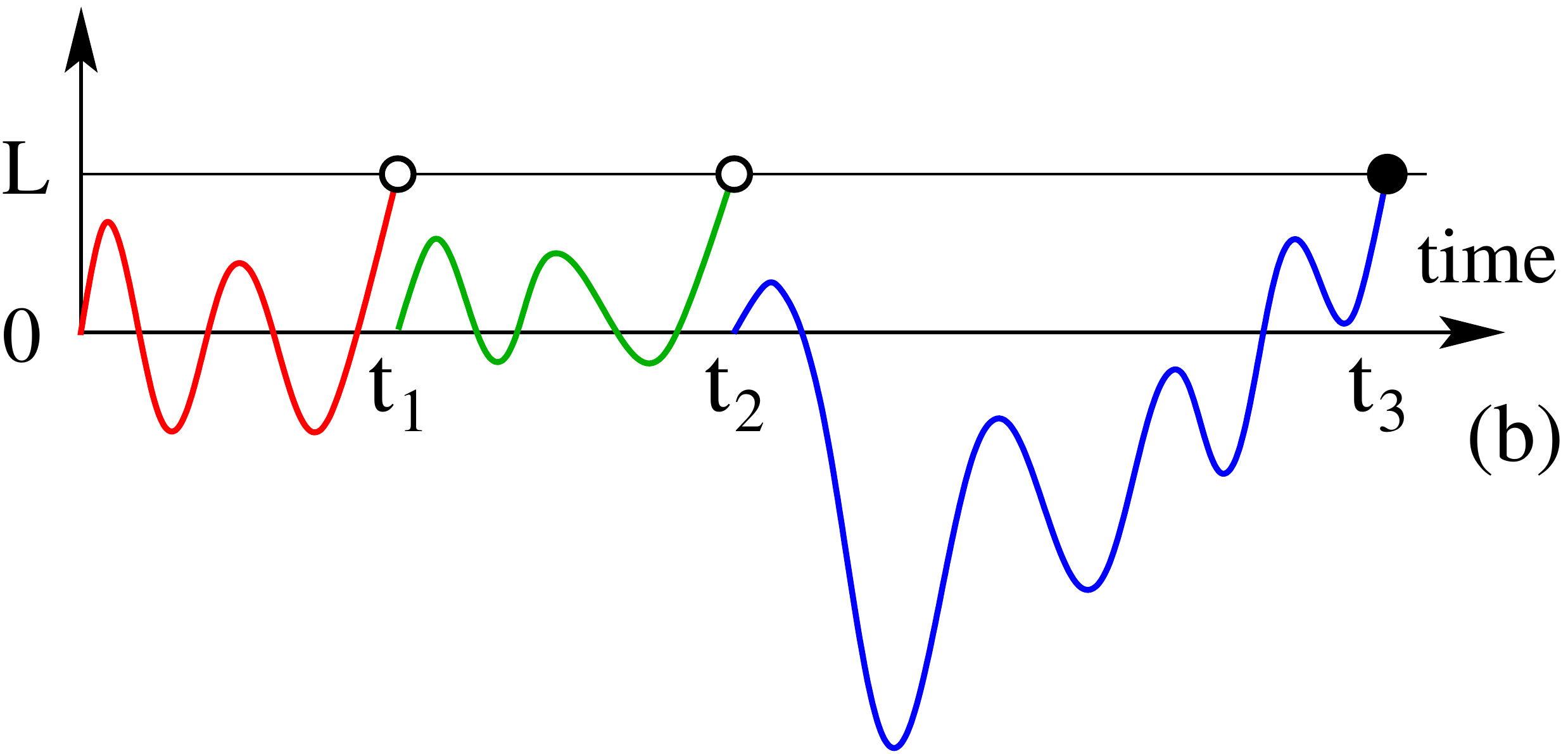}}}
\caption{Relation between a first-passage path to $x=3L$ and a
  third-passage path to $x=L$, with resetting each time $x=L$ is reached.
  The green and blue paths in (b) have merely been shifted vertically
  downward by $L$ and $2L$ compared to (a), respectively.}
\label{fig:reln}
\end{figure}

In hindsight, this equivalence between the first-passage probability to
$x=nL$ and the $n^{\text th}$-passage probability to $x=L$ with resetting at
$x=L$ is self evident.  As indicated in Fig.~\ref{fig:reln} for the case
$n=3$, a first-passage path from 0 to $3L$ is composed of a first-passage
path from 0 to $L$, followed by a first-passage path from $L$ to $2L$, and
finally a first-passage path from $2L$ to $3L$.  Resetting causes each of
these three segments to (re)start from the origin.  Thus the point $x=L$ is first
reached for the third time after resetting by these displaced paths.

\subsection{Spatial probability distribution}

We now compute the probability distribution of the diffusing particle at time
$t$, $P(x,t)$, on the semi-infinite line $x\leq L$ under the influence of
first-passage resetting.  This distribution can be obtained in several ways.
Here we make use of the path transformation discussed above for the
derivation of the $n^{\rm th}$ passage probability~(see Fig.~\ref{fig:reln}).
(A calculational approach based on Laplace transforms that relies on the
renewal structure of the process is given in \ref{app-a}.)
When the walker is at position $x \in [0,L]$ at time $t$ and has experienced
exactly $n$ resets, this is equivalent to a free particle being at position
$x+nL$ without having reached level $(n+1)L$.  As a result, the corresponding
probability is that of a free particle with position $x(t)=x+nL$ and running
maximum position $M(t)<(n+1)L$; the latter is defined by
$M(t)=\max_{t' \leq t}x(t')$. The joint distribution of the position and
maximum, $x(t)=x$ and $M(t)=m$, is given by
\begin{equation}
\label{Pixmt}
\Pi(x,m,t) = \frac{2m-x}{\sqrt{4\pi D^3 t^3}} \,\,e^{-(2m-x)^2/4Dt}\,.
\end{equation}
This formula was established by L\'evy~\cite{Levy1948,bosa12} and it can be
derived by using the reflection property of Brownian motion. From this joint
probability, we find that
\begin{align}
\label{Pxt>}
  P(x,t) &= \sum_{n\geq 0} \mathrm{Prob}\big(M(t)<(n+1)L\text{ and } x(t) = x+nL\big)\nonumber\\
  &= \sum_{n\geq 0}\,\int_{x+nL}^{(n+1)L}\,dm\, \Pi(x+nL,m,t)\nonumber\\
  &=  \frac{1}{\sqrt{4\pi Dt}}\,\sum_{n\geq 0}\left[ e^{-(x+nL)^2/4Dt}
   - e^{-[x-(n+2)L]^2/4Dt}\right],\qquad 0<x\leq L\,.
\end{align}
While this expression is exact, it is not in a convenient form to determine
its long-time behavior.  However, the long-time limit of $P(x,t)$ is simple
to obtain by expanding its Laplace transform (see~\cite{bruyne2020firstpassage}) for small $s$,
\begin{subequations}
\begin{align}
  \widetilde{P}(x,s)\simeq \frac{1}{\sqrt{Ds}}\, \frac{L-x}{L}\qquad 0\leq
  x\leq L\, ,\quad s\rightarrow 0\,,
\end{align}
from which the inverse Laplace transform is
\begin{align}
  P(x,t)\simeq \frac{1}{\sqrt{\pi Dt}}\, \frac{L-x}{L}\qquad 0\leq
  x\leq L\, ,\quad t\rightarrow \infty\,.
\end{align}
\end{subequations}
The linear $x$ dependence arises from the balance between the diffusive
flux that exits at the reset point $x=L$ and this same flux being re-injected
at $x=0$.

For $x<0$, the integral in the second line of \eqref{Pxt>} ranges from $nL$ to $(n+1)L$ rather than from $x+nL$. This leads to
\begin{align}
\label{Pxt<}
  P(x,t)  &=  \frac{1}{\sqrt{4\pi Dt}}\,\sum_{n\geq 0}\left[ e^{-(x-nL)^2/4Dt}
   - e^{-[x-(n+2)L]^2/4Dt}\right]\nonumber\\
   &=  \frac{1}{\sqrt{4\pi Dt}}\,
  \left[e^{-x^2/4Dt}+ e^{-(x-L)^2/4Dt}\right],\qquad x<0\,.
\end{align}
Thus the probability distribution is merely the sum of two Gaussians.  In the
long-time limit and for $|x|/\sqrt{4Dt}\gg 1$, the factor $L$ in the
second term becomes irrelevant and the distribution reduces to that of
diffusion on the half line in the presence of a reflecting boundary.

\begin{figure}[ht]
\centerline{
\includegraphics[width=0.45\textwidth]{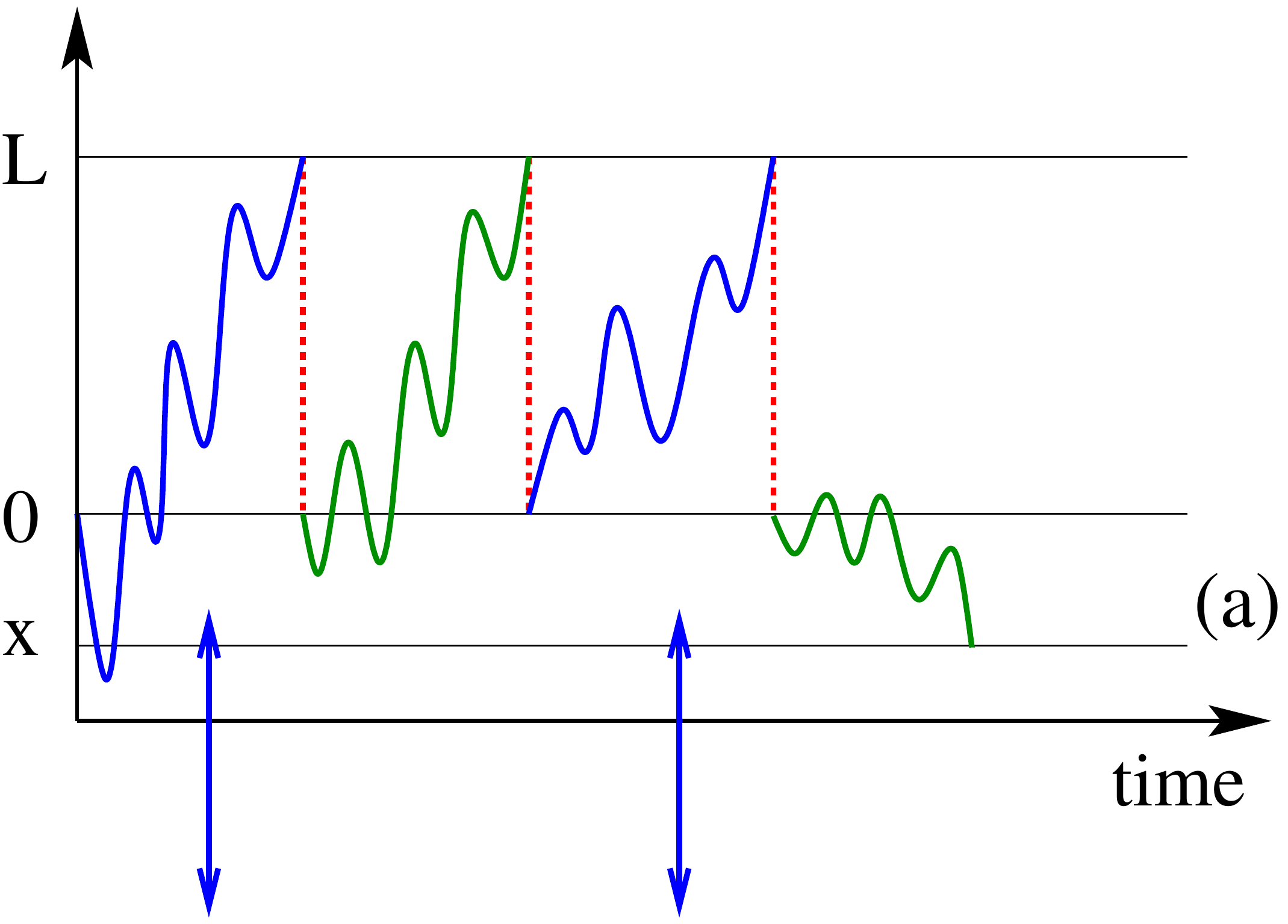}\quad
\includegraphics[width=0.45\textwidth]{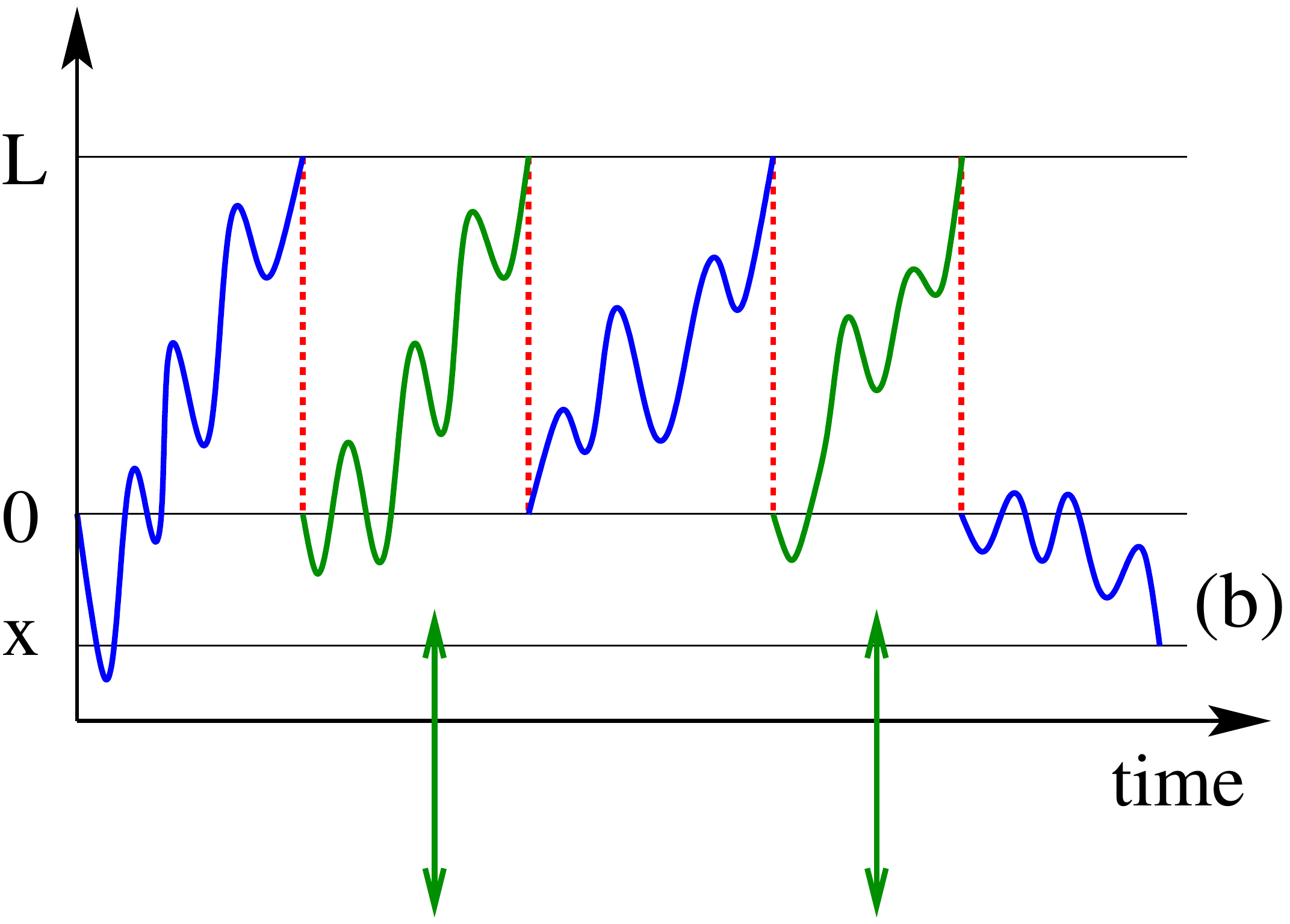}}
\vspace*{-10mm}
\centerline{
\includegraphics[width=0.45\textwidth]{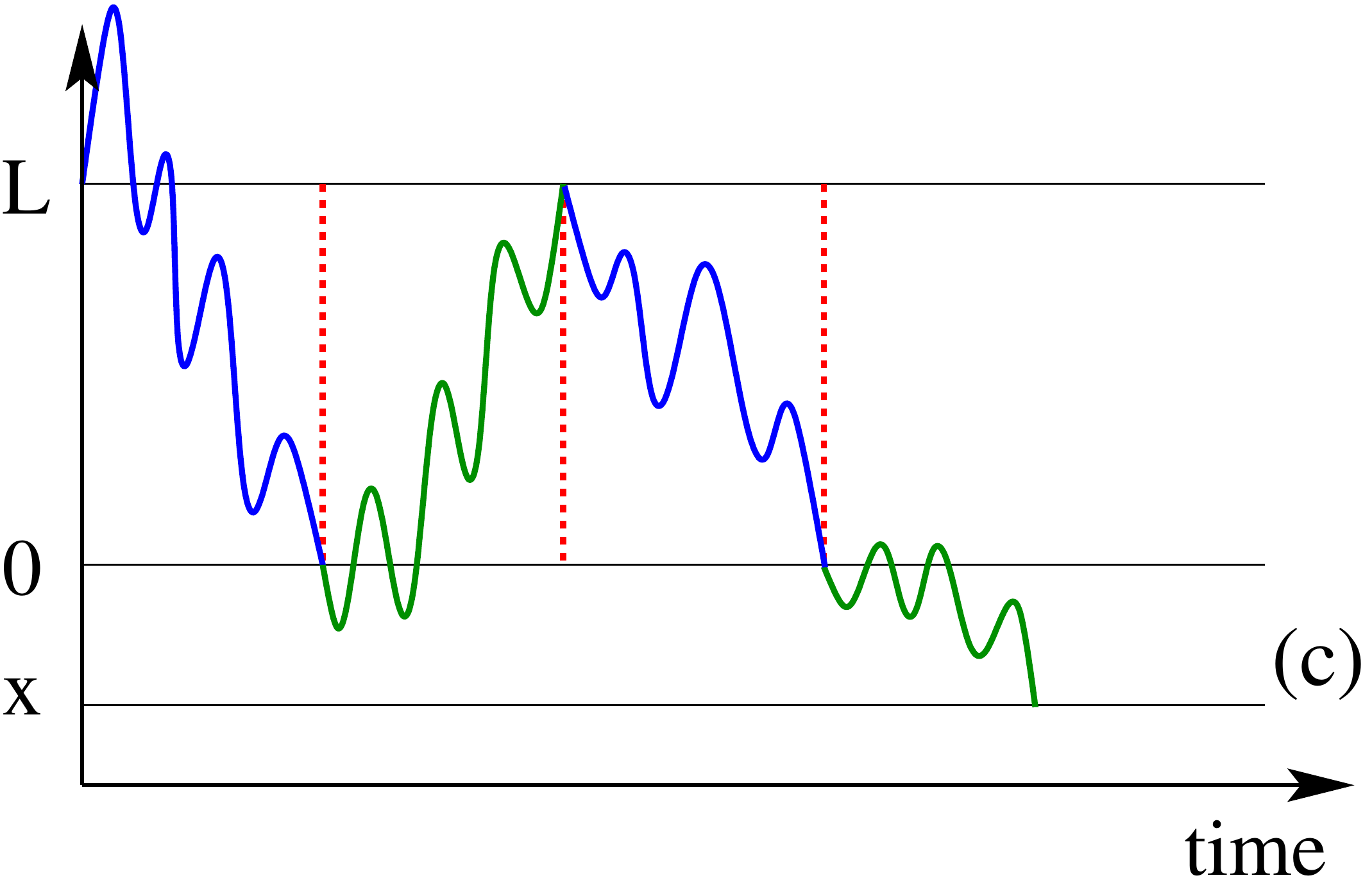}\quad
\includegraphics[width=0.45\textwidth]{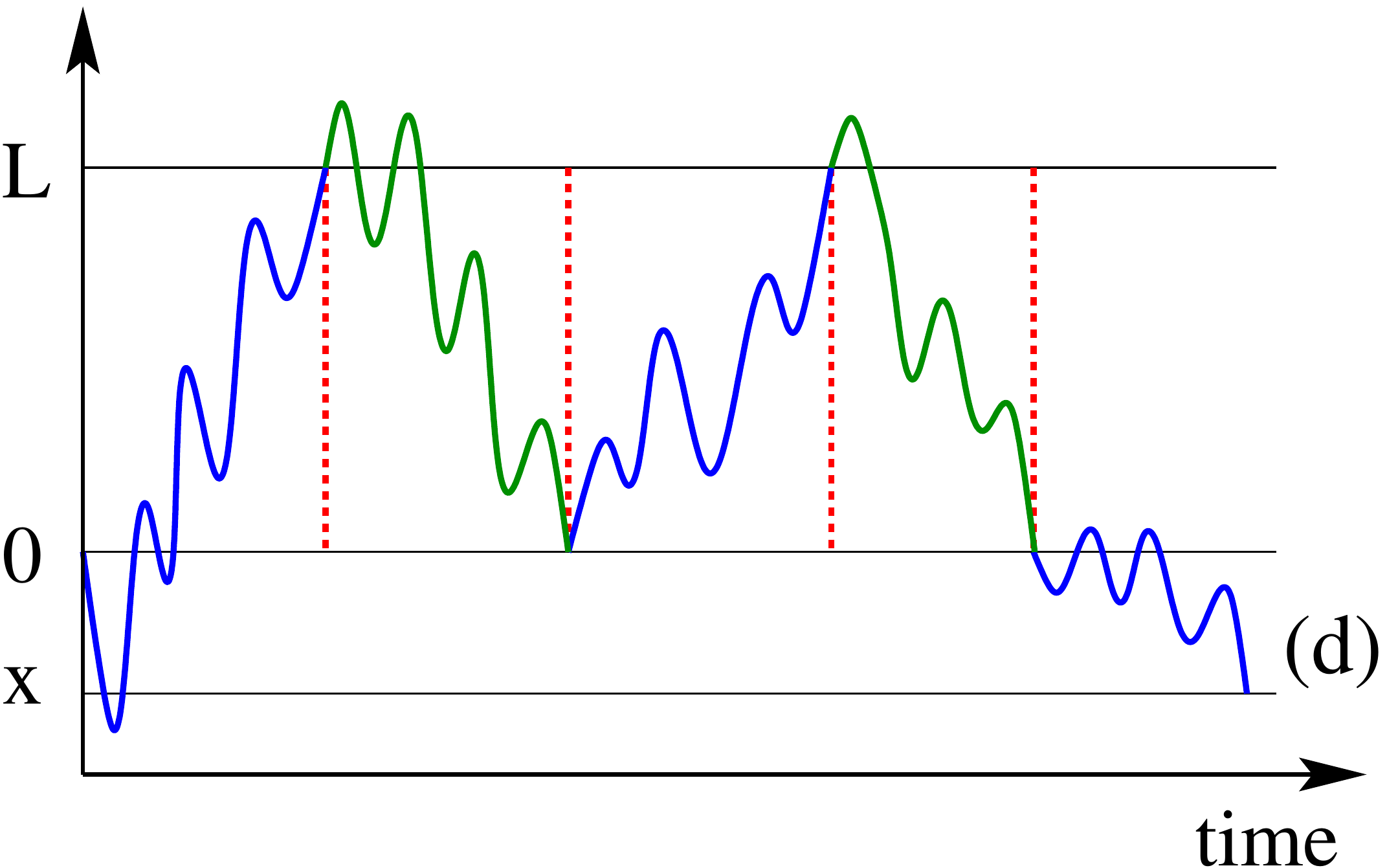}}
\vspace*{2mm}
\caption{Schematic space-time trajectory of diffusion with first-passage
  resetting on a semi-infinite line.  (a) A path with an odd number of resets
  is equivalent to (c) a freely diffusing path that starts at $x(t\!=\!0)=L$.
  (b) A path with an even number of resets is equivalent to (d) a freely
  diffusing path that starts from $x(t\!=\!0)=0$. This equivalence underlies
  the spatial probability distribution for $x<0$ in Eq.~\eqref{Pxt<}.}
    \label{fig:pathtsfm}
\end{figure}

An appealing way to obtain the probability distribution \eqref{Pxt<} is by a
path decomposition construction.  Consider the original resetting problem and
partition all trajectories into those that undergo either an odd or an even
number of resets.  In the former case, we invert the segments \emph{before}
each odd-numbered resetting about the origin and then translate each such
segment by a distance $+L$ (blue arrows in Fig.~\ref{fig:pathtsfm}(a) \&
(c)).  As shown in this portion of the figure, the resulting trajectory is
simply a Brownian path that starts at $x=L$ and propagates freely to its
final position $x$.  For a path that consists of an even number of resetting
events, we perform this same inversion and translation on the segments
\emph{after} each odd-numbered resetting (green arrows in
Fig.~\ref{fig:pathtsfm}(b) \& (d)).  The resulting trajectory is now a
Brownian path that starts at the origin and propagates freely to its final
position $x$.  It is worth emphasizing that this decomposition applies for
any symmetric and continuous stochastic process (provided it is homogeneous and stationary).

\subsection{Average number of resets}
\label{subsec:nav}

To find the average number of resets that occur up to time $t$, we first
compute the probability that $n$ resets have occurred during this time.  By
relying on the path transformation shown in Fig.~\ref{fig:reln}, we find that
the probability for exactly $n$ resets to occur by time $t$ equals the
probability for a freely diffusing particle to have a running maximum $M(t)$
greater than $nL$ but less than $(n+1)L$, that is:
\begin{align}
  P(N(t)=n) &= \mathrm{Prob}\big(nL\leq M(t)<(n+1)L\big)\,.
\end{align}
The distribution of $M(t)$ is
known~\cite{bachelier1900theorie,levy1940certains,bosa12} and may be readily
rederived from~\eqref{Pixmt},
\begin{align*}
  P(M(t)=m) =\frac{1}{\sqrt{\pi D t}}\,\, e^{-m^2/4Dt}\,,
\end{align*}
from which it follows that
\begin{equation}
\label{eq:ndist}
P\big(N(t)\!=\!n\big)=\mathrm{erf}\left(\frac{(n+1)L}{\sqrt{4Dt}}\right)-\mathrm{erf}\left(\frac{nL}{\sqrt{4Dt}}\right),
\end{equation}
where $\mathrm{erf}$ is the Gauss error function.

We can compute the average number of reset events,
$\mathcal{N}(t)\equiv \left\langle N(t) \right\rangle$,
from~\eqref{eq:ndist}, but it is quicker to use again the mapping with the
running maximum of free diffusion.  Indeed, writing $\mathcal{M}(t)$ for the
average maximum position of a freely diffusing particle up to time~$t$, one
has
\begin{equation}
\mathcal{N}(t)\,L\, \leq\, \mathcal{M}(t) \, < \, \left[\mathcal{N}(t)+1\right]\,L.
\end{equation}
Since $\mathcal{M}(t)=\sqrt{4Dt/\pi}$, we find that the long-time behavior of $\mathcal{N}(t)$ is given by
\begin{align}
\label{Nt}
  \mathcal{N}(t) \simeq \sqrt{4Dt/\pi L^2}\,.
\end{align}
The resetting process is non-stationary, as the number of reset
events grows as~$\sqrt{t}$.

\subsection{Biased diffusion}
\label{subsec:bias}

The case where the diffusing particle is biased towards the resetting
boundary is equivalent to the model studied by Falcao and
Evans~\cite{falcao2017interacting}.  Here we briefly discuss the
complementary situation in which the particle is biased away from the
resetting boundary, with drift velocity $v<0$.  In the absence of resetting,
a particle that starts at the origin eventually reaches $x=L$ with
probability $H=e^{-\Pe}$ and escapes to $x\!=\!-\infty$ with probability
$1-H$~\cite{feller2008introduction,redner2001guide}, where $\Pe\equiv vL/2D$
is the P\'eclet number (the dimensionless bias velocity).  When resetting can
occur, $H$ now becomes the probability that a resetting event actually
happens.  Consequently, the probability that the particle resets exactly $n$
times is given by $R_n=H^n(1-H)$.  The average number of resetting events
before ultimate escape therefore is~\cite{Levy1948}
\begin{align}
  \mathcal{N}(t) = \sum_n n R_n = \frac{H}{1-H} =
  \frac{1}{e^{\Pe}-1}\,.
\end{align}
In the limit of $v\to 0$, the number of resetting events diverges as
$\mathcal{N}(t) \simeq 2D/(vL)$.  The time between resetting events is
known to be $L/v$~\cite{redner2001guide,krapivsky2018first}.  Thus after
typically $1/(e^{\Pe}-1)$ resetting events, each of which requires a time of
$L/v$, the particle escapes to $-\infty$.

We can also compute the spatial probability distribution of the particle when
it undergoes biased diffusion with bias velocity of magnitude $v$. We make
use again of the path transformation shown in Fig.~\ref{fig:reln}. For a
Brownian particle with drift $v$, the analog of Eq.~\eqref{Pixmt} is
\begin{equation}
\label{PixmtDr}
\Pi(x,m,t) = \frac{2m-x}{\sqrt{4\pi D^3 t^3}} \,\,e^{-(2m-x)^2/4Dt}\,
e^{\Pe\,\left(x/L-\Pe\,Dt/L^2\right)}\,,
\end{equation}
which leads to
\begin{subequations}
\begin{align}
\label{Pxtdr>}
  P(x,t) &=  \frac{1}{\sqrt{4\pi Dt}}\,\sum_{n\geq 0}\left[ e^{-(x+nL)^2/4Dt}
   - e^{-[x-(n+2)L]^2/4Dt}\right]\,e^{\Pe\,\left(x/L+n-\Pe\,Dt/L^2\right)},\quad 0<x\leq L\,
\end{align}
and
\begin{align}
\label{Pxtdr<}
  P(x,t)  &=  \frac{1}{\sqrt{4\pi Dt}}\,\sum_{n\geq 0}\left[ e^{-(x-nL)^2/4Dt}
   - e^{-[x-(n+2)L]^2/4Dt}\right]\,e^{\Pe\,\left(x/L+n-\Pe\,Dt/L^2\right)},\quad x<0\,.
\end{align}
\end{subequations}
In contrast to the driftless case, there is no simplification for the
probability distribution when $x<0$. In particular, the path transformation
of Fig.~\ref{fig:pathtsfm} requires symmetry and thus does not hold in the
presence of drift.

\section{Optimization in First-Passage Resetting}
\label{sec:finiteInt}

\subsection{The finite interval}
\label{subsec:nodelay}

We now introduce an optimization problem that is induced by first-passage
resetting.  We envisage a mechanical system whose operating coordinate $x(t)$
lies in the range $[0,L]$.  Increasing the value of $x$ corresponds to
increasing its level of operation, and it is desirable to run the system as
close as possible to its maximum capacity $L$.  However, the system breaks
down whenever $x$ reaches $L$, after which repairs have to be made before the
system can restart its operation from $x=0$.  While the dynamics of the
operating coordinate is typically complicated and dependent on multiple
parameters, we view the coordinate $x$ has undergoing a drift-diffusion
process for the sake of parsimonious modeling.  For the system to be close to
$x=L$, the drift should be positive.  On the other hand, breakdowns of the
system are to be avoided because a cost is incurred with each breakdown.
This suggests that the drift velocity should be negative.  The goal is to
determine the optimal operation that maximizes the performance of the system
as a function of the cost for each breakdown and the drift velocity.
Although the analogy between first-passage resetting and a mechanical system
is naive, this formulation allows us to determine the optimal operation in a
concrete way.

The basic control parameter is the magnitude of the drift velocity.  If the
velocity is large and negative, the system is under-exploited because it
operates far from its maximum capacity.  Conversely, if the velocity is large
and positive, the system breaks down often.  We seek the optimal operation by
maximizing an objective function $\mathcal{F}$ that rewards high performance
and penalizes breakdowns.  A natural choice for $\mathcal{F}$ is
\begin{align}
\label{F}
\mathcal{F} =\lim_{T\to\infty} \frac{1}{T}\left[ \frac{1}{L} \int_0^T x(t)\,dt - C\, \mathcal{N}(T)\right]\,,
\end{align}
where $T$ is the total operation time, $\mathcal{N}(T)$ is the average number
of breakdowns within a time $T$, with $T$  much longer than the mean breakdown
time, and $C$ is the cost of each breakdown.  As defined, this objective
function rewards operation close to the maximum point $L$ and penalizes
breakdowns.

We now determine this objective function when the operating coordinate $x(t)$
evolves according to drift-diffusion, with the additional constraint that
$x(t)$ is reset to zero whenever $x$ reaches $L$.  Mathematically, we need to
solve the convection-diffusion equation with the following additional
conditions: (i) a $\delta$-function source at the origin whose magnitude is
determined by the outgoing flux $j(x)=-D\partial_x c+vc$ at $x=L$, (ii)
a reflecting boundary condition at $x=0$, and (iii) the initial condition
$x(t=0)=0$.  That is, we want to solve
\begin{subequations}
\begin{align}
\label{eq:difft}
  \partial_t c  +v\partial_x c
  =D \partial_{xx}c + \delta(x)(-D\partial_x c+vc)\big|_{x=L} \,,
 \end{align}
subject to 
\begin{align*}
\begin{cases}
    \left(D\partial_x c-vc\right)\rvert_{x=0} = \delta(t) \,&\\
    c(L,t) = 0 \,&\\
    c(x,0) = 0\,&
    \end{cases}.
\end{align*}
Here, $c\equiv c(x,t)$ is the probability density for the operating
coordinate, the subscripts denote partial differentiation, $D$ is the
diffusion coefficient, and $v$ is the drift velocity.  Notice that the
reflecting boundary condition at $x=0$ holds \emph{except} at the start of
the process to account for the unit input of flux at $t=0$.  This
construction allows one to take the initial condition to be $c(x,t\!=\!0)=0$,
which greatly simplifies all calculations.  Effectively, this flux initial
condition corresponds to starting the system with the particle at $x=0$.

As in Sec.~\ref{sec:semiInfiniteGeometry}, we first solve the free theory,
where the delta-function term in \eqref{eq:difft} is absent, and then use
renewal equations to solve the full problem.  In the free case,
Eq.~\eqref{eq:difft} becomes, in the Laplace domain:
\begin{align}
\label{eq:difffree}
  s\widetilde{c_0}+ v\partial_x \widetilde{c_0}
  =D \partial_{xx}\widetilde{c_0} \,,
\end{align}
and is subject to the boundary conditions
\begin{align*}
\begin{cases}
    \left(D\partial_x \widetilde{c_0}-v\widetilde{c_0}\right)\rvert_{x=0} = 1\, \\
    \widetilde{c_0}(L,s) = 0 \,,
    \end{cases}
\end{align*}
\end{subequations}
where the subscript $0$ denotes the concentration without flux re-injection.
The solution to \eqref{eq:difffree} is standard and the result is (see
\ref{app:applybc} for details):
\begin{subequations}
  \label{eq:c0F1}
\begin{align}
\label{eq:c0}
  \widetilde{c}_0(x,s) &= \frac{2\, e^{\P} \sinh \left[w (L-x)\right]}{\mathcal{W}}\,,
\end{align}
where $\P\equiv vx/2D$, $w = \sqrt{v^2+4Ds}/2D$ and
$\mathcal{W}= 2 D w \cosh (L w)+v \sinh (L w)$.  In terms of
$\widetilde{c}_0$, the Laplace transform of the first-passage probability to
$x=L$ is
\begin{align}
\label{eq:F1}
\widetilde{F}_1(L,s) &= \left( -D\partial_x \widetilde{c}_0 + v \widetilde{c}_0 \right)\rvert_{x=L}
 = \frac{2 \,D\, w\, e^{\Pe}}{\mathcal{W}}\,,
\end{align}
\end{subequations}
where again $\Pe = vL/2D$ is the P\'eclet number. With re-injection of the
outgoing flux, the concentration obeys the renewal equations. In the Laplace
domain and using $\widetilde{c}_0$ above from the free theory, we obtain
\begin{align}
\label{eq:part}
  \widetilde{c}(x,s)= \frac{\widetilde{c}_0(x,s)}{1-F_1(L,s)}
  = \frac{2\, e^{\P} \sinh [w (L-x)]}{\mathcal{W}-2 D\, w \,e^{\Pe}}\,,
\end{align}
where we substitute in the results from Eqs.~\eqref{eq:c0F1} to obtain the final result.
\smallskip

\begin{figure}[ht]
    \centering
    \includegraphics[width=0.7\textwidth]{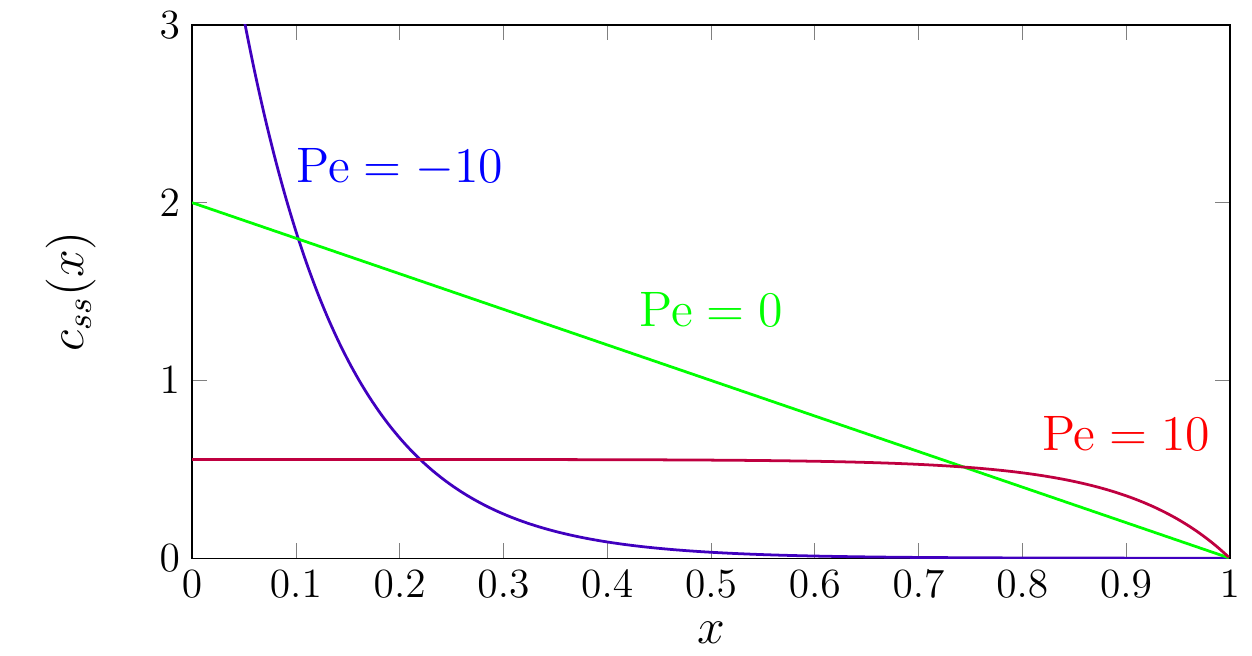}
    \caption{The stationary distribution for the first-passage resetting
      process on the interval $[0,1]$ for different P\'eclet numbers.
    }
    \label{fig:cavg}
\end{figure}

Contrary to the semi-infinite case, a stationary distribution is attained on
the finite interval.  To determine this steady state, we use the duality
between the limits $s\to 0$ in the Laplace domain and $t\to \infty$ in the
time domain.  With this approach, the coefficient of the term proportional to
$1/s$ in $\widetilde{c}(x,s)$ gives the steady-state concentration,
$c_{\text{ss}}$, in the time domain:
\begin{align}
  \label{eq:solt}
  c_{\text{ss}}(x)  &\simeq
    \frac{1}{L}\times\frac{1\,-\,e^{-2(\Pe-\P)}}
      {1\,-\,\Pe^{-1}\,e^{-\Pe}\,\sinh\left(\Pe\right)}\,,
\end{align}
from which the normalized first moment in the steady state is
\begin{align}
\label{eq:xavg}
\frac{\langle x\rangle}{L} =\frac{1}{L}\int_0^L \!\!x\,c(x)\, dx=
 \frac{\left(2\Pe^2-2\Pe+1\right)\,e^{2\Pe}-1}{ 2\Pe\left[(2\Pe-1)\,e^{2\Pe}+1\right]} \,.
\end{align}
Representative plots of the stationary-state concentration for different
P\'eclet numbers are given in Fig.~\ref{fig:cavg}.  As one might anticipate,
the density profile is concentrated near $x=0$ for negative drift velocity,
while for positive drift there is a constant cycling of outgoing flux that is
reinjected at $x=0$, which leads to a nearly constant density profile.

\begin{figure}[ht]
     \centerline{
       \subfigure[]{\includegraphics[width=0.475\textwidth]{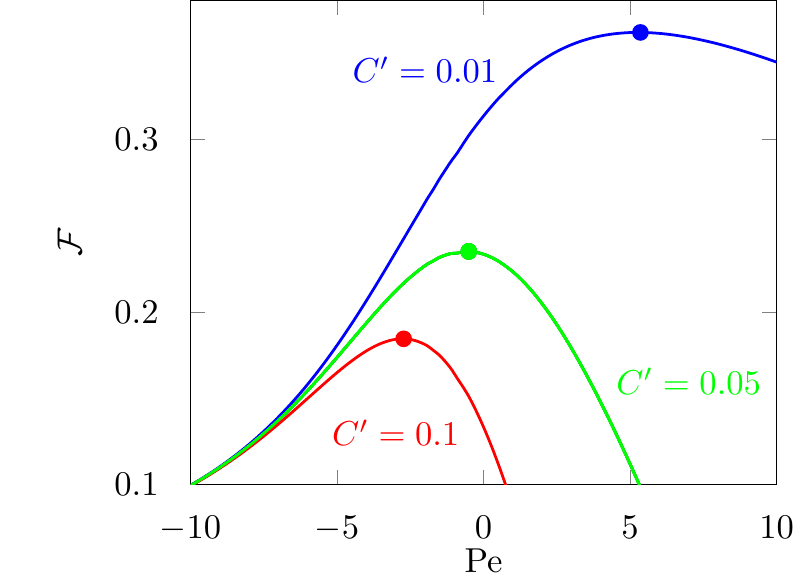}}\quad
\subfigure[]{\includegraphics[width=0.475\textwidth]{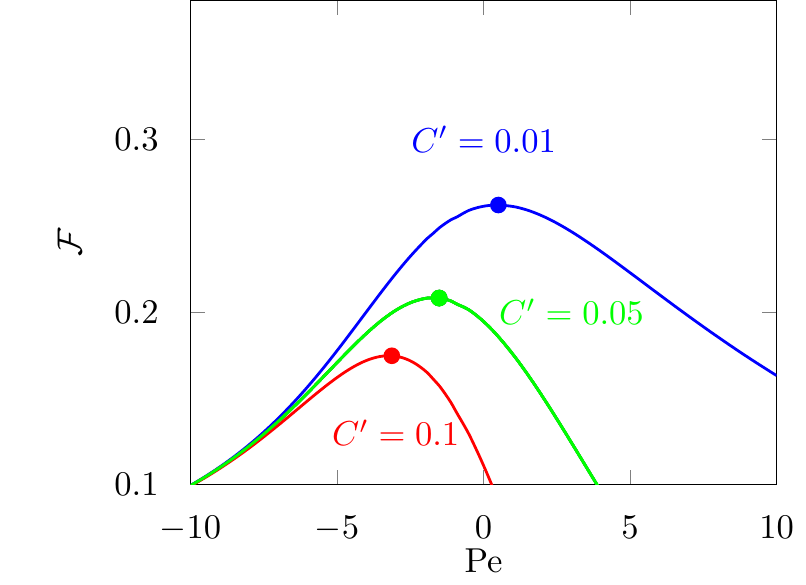}}}
\caption{The objective function versus P\'eclet number $\Pe$ for different
  normalized cost values $C'\equiv C/(L^2/D)$ for: (a) no delay upon
  resetting, and (b) a dimensionless delay time of $\overline{\tau}=0.1$ at
  each resetting.  Indicated on each curve is the optimal operating point. }
     \label{fig:F}
\end{figure}

The average number of reset events $\mathcal{N}$ satisfies the renewal
equation~\eqref{eq:navgb}, and substituting in $\widetilde{F}_1$ from
\eqref{eq:F1}, we obtain
\begin{subequations}
\begin{align}
\label{eq:Nssp}
\widetilde{\mathcal{N}}(s) &=  \frac{2 D\, w\, e^{\Pe}}  {s\,
 \left[\mathcal{W}-2 D w\, e^{\Pe}\right]}\,.
\end{align}
We now extract the long-time behavior for the average number of times that
$x=L$ is reached by taking the limit $s\rightarrow 0$ of
$\widetilde{\mathcal{N}}(s)$ to give
\begin{align}
\label{eq:N}
\mathcal{N}(T) &\simeq \frac{4\Pe^2}{2\Pe-1+ e^{-2\Pe}}\,\,\frac{T}{L^2/D}\,.
\end{align}
\end{subequations}
Substituting these expressions for $\langle x\rangle/L$ and $\mathcal{N}$
into \eqref{F} immediately gives the objective function, and representative plots
are shown in Fig.~\ref{fig:F}(a).  For a given cost of a breakdown, there is
an optimal drift velocity or optimal P\'eclet number.  The higher this cost,
the smaller the optimal bias and the value of $\mathcal{F}$.  Moreover, the
optimal bias is not necessarily negative. Indeed, if the cost of a breakdown
is relatively small, then it is advantageous to operate the system close to
its limit $L$ and absorb the (small) cost of many breakdowns.  On the
contrary, if the cost of a breakdown is high, it is better to run the system
at low level and with a negative bias to avoid breakdowns.

\subsection{Time delay for repair}
\label{sec:delay}

When a mechanical system breaks down, there is usually some downtime during
which repairs are made before the system can be restarted.  Such a downtime
can naturally be incorporated into our model by including a random delay time
after each resetting event.  Thus when the particle reaches $x=L$ and is
returned to $x =0$, we posit that the particle waits at the origin for a
random time $\tau$ that is drawn from the exponential distribution
$\sigma^{-1}e^{-\tau/\sigma}$ before the particle starts moving again.  We
now determine the role of this delay on the optimal operation of the system.

The governing renewal equations can be readily extended to incorporate this
delay.  This delay mechanism can also be viewed as the so-called ``sticky''
Brownian
motion~\cite{gallavotti1972boundary,harrison1981sticky,bou2020sticky} that is
then combined with first-passage resetting.  When we include this delay, the
renewal equation for the probability distribution becomes:
\begin{subequations}
  \label{eq:delayRen}
\begin{align}
  \label{eq:renEqT}
P(x,t) = G(x,L,t) + \int_0^t \!\!dt' &F_1(t')\bigg[ \delta_0(x) e^{- (t-t')/\sigma} 
+   \int_0^{t-t'}\!\!\frac{d\tau}{\sigma}\, e^{- \tau/\sigma} P(x,t\!-\!t'\!-\!\tau)\bigg].
\end{align}
This equation encapsulates the two possibilities for the subsequent behavior
of the particle when it first reaches $x=L$ at time $t'$.  Either the
particle remains at $x=0$ for the remaining time $t-t'$ or the particle waits
for a time $\tau<t-t'$ and then the process starts anew from
$(x,t)=(0,t'+\tau)$ for the remaining time $t-t'-\tau$.

In a similar fashion, the renewal equation for the average
number of resetting events is
\begin{align}
  \label{eq:navgbT}
    \begin{split}
  \mathcal{N}(t) = \int_0^t dt' &F_1(t')\bigg\{ e^{- (t-t')/\sigma}
  + \int_0^{t-t'}\frac{d\tau}{\sigma}
  e^{- \tau/\sigma} \big[1 +\mathcal{N}(t-t'-\tau)\big]\bigg\}.
                                   \end{split}
\end{align}
\end{subequations}
Equation~\eqref{eq:navgbT} accounts for the particle first hitting $L$ at
time $t'$ and either waiting at the origin for the entire remaining time
$t-t'$ or waiting there for a time $\tau<t-t'$ and then renewing the process
for the remaining time.  For this latter possibility, there will be, on
average, $1 +\mathcal{N}(t-t'-\tau)$ resetting events. 

Solving Eqs.~\eqref{eq:delayRen} in the Laplace domain yields:
\begin{subequations}
\begin{align}
  \label{PN}
\begin{split}
  \widetilde{P}(x,s) &= \frac{\delta(x)\, \sigma \widetilde{F}_1(s) +
  \widetilde{G}(x,s)(1+\sigma s)}{1-F_1(s) +\sigma s}\,, \\
  \widetilde{\mathcal{N}}(s) &= \frac{\widetilde{F}_1(s)(\sigma
    +1/s)}{1+s\sigma -\widetilde{F}_1(s)}\,.
\end{split}
\end{align}
We now use the results from Sec.~\ref{subsec:nodelay} for the optimization
problem on the interval with no delay.  Namely, we substitute in
Eqs.~\eqref{PN} the first-passage probability $\widetilde{F}_1(s)$ from
Eq.~\eqref{eq:F1} and the probability distribution in Eq.~\eqref{eq:c0} for
$\widetilde{G}(x,s)$ to obtain
\begin{align}
\begin{split}
  \widetilde{P}(x,s) &= \frac{\delta(x)\, \sigma 2D w +
                        2\sinh(w(L-x))(1+\sigma
                        s)}{\mathcal{W}e^{-\Pe}(1+\sigma s)-2Dw}, \\
  \widetilde{\mathcal{N}}(s) &= \frac{2Dw(\sigma
    +1/s)}{\mathcal{W}e^{-\Pe}(1+s\sigma) -2Dw}\,.
\end{split}
\end{align}
\end{subequations}

From the Laplace transform of the spatial probability density, we compute its
stationary distribution by taking the $s\rightarrow 0$ limit and obtain
\begin{subequations}
\begin{align}
  P(x) \simeq \frac{e^{\Pe}\, \Pe \, (2 \overline{\tau}\, \Pe\, L\,  \delta_0 (x)+1)-\Pe\,
   e^{\P}}{ e^{\Pe} \left(\overline{\tau}\,
   \Pe^2+\Pe-1\right)+1}\frac{1}{L},
\end{align}
where $\overline{\tau}=D\sigma /L^2$ is the dimensionless delay time.  From
this distribution, the average position of the particle is
\begin{align}
\label{eq:xav-d}
\frac{\langle x\rangle}{L}=  \frac{\left[(\Pe-2)\Pe+2\right]e^{\Pe}-2}
  {2 \left[\Pe(\overline{\tau}\,\Pe^2+\Pe-1)e^{\Pe}+\Pe\right]} \,.
\end{align}
\end{subequations}
Similarly, the average number of resetting events, or equivalently, the
average number of breakdowns in the long-time limit is
\begin{align}
  \mathcal{N} =\frac{\Pe^2}{\Pe-1+\overline{\tau}\,\Pe^2+e^{-\Pe}}\,\frac{T}{L^2/D}\, .
\end{align}
These two results, when substituted into Eq.~\eqref{F}, give an objective
function $\mathcal{F}$ whose qualitative features are similar to the case of
no delay (Fig.~\ref{fig:F}(b)).  This behavior is what might anticipate,
since delay may be viewed as an additional form of cost.

The primary difference with the no-delay case is that the optimal P\'eclet
number and the corresponding optimal objective function $\mathcal{F}$ both
decrease as the delay time is increased (Fig.~\ref{fig:F}).  Indeed, delay
reduces the number of resetting events/breakdowns, but also induces the
coordinate to remain closer to the origin.  In the limit where the delay is
extremely long, the optimal P\'eclet number will be small.  Moreover this
optimal value will be nearly independent of the cost per breakdown, as the
particle will almost never hit the resetting boundary.

\subsection{Two dimensions}
\label{sec:multiCompnt}

It is natural to extend the optimization problem on the interval to
higher-dimensional domains.  Here, we treat the case where the domain is an
annulus of outer radius $L$, inner radius $a<L$, and the diffusing particle
is reset to $r=a$ whenever the outer domain boundary is reached.  In analogy
with the one-dimensional problem, the particle also experiences drift
velocity $v(r)=v_0/r$.  As we shall see, the choice of a potential flow field
is convenient because the velocity can be combined with the centrifugal term
in the Laplacian, which simplifies the form of the solution.  The finite
inner radius is needed to eliminate the infinite-velocity singularity that
would occur if the inner radius was zero.

In close analogy with the finite-interval system, Eq.~\eqref{eq:difft}, the
equation of motion for the particle is
\begin{align}
  \label{eq:c2d}
  \partial_t c + \frac{v}{r}\partial_rc
  = D\left(\partial_{rr}c +\frac{1}{r}\partial_r c\right)
  + \delta_a(r)\big[2\pi r(-D\partial_r c + vc)\big]\big|_{r=L}\, .
\end{align}
Here, the flux term has a factor $2\pi r$ due to an integration over all
angles.  In this geometry, the probability density of finding a particle at a
radius $r$ is $2\pi r\, c(r,t)$.  We now introduce the dimensionless variables
$x=r/L$, $x_0=a/L$, the Fourier number $\Fo=Dt/L^2$, and
$\Pe=v_0/D$\footnote{Note that $v_0$ has units of velocity times length, so
  this definition of the P\'eclet number is dimensionally correct.}  to
transform Eq.~\eqref{eq:c2d} into
\begin{align}
\label{eq:c2da}
 \partial_\Fo c(x,\Fo) &=  \partial_{xx}c(x,\Fo) + \frac{1-\Pe}{x}\,\partial_x  c(x,\Fo)
    + \delta_{x_0}(x)\big[2\pi x(-\partial_x c + \Pe\, c)\big]\big\rvert_{x=1}, 
\end{align}
and the appropriate boundary conditions for this equation are
\begin{align*}
\begin{cases}
  \big[\Pe\, c(x,\Fo)- x \partial_x c(x,\Fo)\big]\big\rvert_{x=x_0} =\delta(\Fo)/(2\pi)\\
    c(1,\Fo) = 0 \\
    c(x,0) = 0 \,.
\end{cases}
\end{align*}

By performing similar calculations as in the one-dimensional case we find the
following expression for the steady-state probability density in the time
domain (see \ref{app:fpr2d} for the details):
\begin{align}
  \label{eq:solt2d}
  2\pi x\, c(x) &\simeq \frac{2 (\Pe+2) x \left(x^{\Pe}-1\right)}{\Pe \left(x_0^2-1\right)-2
   x_0^2 \left(x_0^{\Pe}-1\right)}\,.
\end{align}
From this expression, the average radial displacement is
\begin{align}
\label{eq:xavg2d}
  \langle x\rangle =\int_{x_0}^1 \!\!x\,\,2\pi\,x\,c(x)\, dx
  =  \frac{2 (\Pe+2) \left[\Pe   \left(x_0^3-1\right)-3 x_0^3
  \left(x_0^{\Pe}-1\right)\right]}{3 (\Pe+3)
  \left[\Pe  \left(x_0^2-1\right)-2 x_0^2   \left(x_0^{\Pe}-1\right)\right]}  \,.
\end{align}

\begin{figure}[ht]
     \centerline{\includegraphics[width=0.475\textwidth]{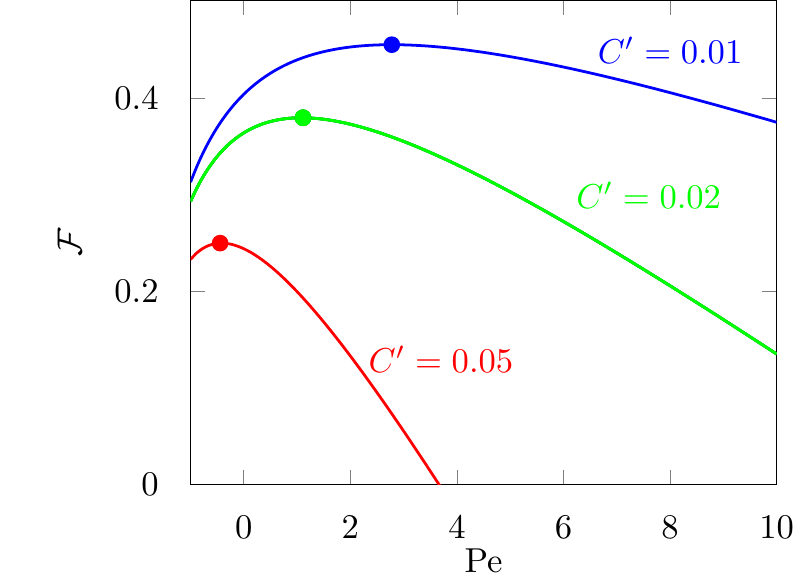}}
\caption{The objective function versus P\'eclet number $\Pe$ for different
  normalized cost values $C'\equiv C/(L^2/D)$ in two dimensions.}
     \label{fig:F2d}
\end{figure}

The average number of reset events $\mathcal{N}$ satisfies a renewal
equation and using $\widetilde{F}_1$ from
Eq.~\eqref{eq:F12d} we find
\begin{subequations}
\begin{align}
\widetilde{\mathcal{N}}(s) &=  \frac{1}{s}\frac{1}{\, \mathcal{W}-1}\,,
\end{align}
where
\begin{align*}
\mathcal{W}=x_0^{1+\Pe/2}  \sqrt{s}\left[
   K_{{\Pe}/{2}}\left(\sqrt{s}\right) I_{1+\Pe/2}\left(\sqrt{s}
   x_0\right)+  I_{{\Pe}/{2}}\left(\sqrt{s}\right)
  K_{1+\Pe/2}\left(\sqrt{s} x_0\right)\right]\,,
\end{align*}
and $I_{\nu}(x)$ and $K_{\nu}(x)$ are the modified Bessel functions of the
first and second kind, respectively.  We now extract the long-time behavior
for the average number of times that $x=L$ is reached by taking the limit
$s\rightarrow 0$ of $\widetilde{\mathcal{N}}(s)$. We find
\begin{align}
\label{eq:N2d}
\mathcal{N}(t) &\simeq\frac{2 \Pe (\Pe+2)}{\Pe+2
   x_0^2   \left(x_0^{\Pe}-1\right)-\Pe\, x_0^2}\,\,\Fo\,.
\end{align}
\end{subequations}

From Eqs.~\eqref{eq:xavg2d} and \eqref{eq:N2d}, we immediately obtain the
objective function and representative results are given in
Fig.~\ref{fig:F2d}.  Overall, the two-dimensional system has the same
qualitative behavior as in one dimension.  In the limit $x_0\rightarrow 0$
the average particle position $\langle x\rangle$ and the average number of
resetting events $\mathcal{N}$ take an even simpler form than in one
dimension:
\begin{align}
     \langle x\rangle &\simeq \frac{2 (\Pe+2)}{3 (\Pe+3)}\,\Theta(\Pe+2)  \,,\\
     \mathcal{N}(T) &\simeq 2(2+\Pe)\,\Theta(\Pe+2)\,\Fo\,,
\end{align}
where $\Theta(x)$ is the Heaviside step function.  The step function arises
because of the curious feature that when $\Pe<-2$, the flow field $\Pe/r$ at
the origin is so strong that the particle remains trapped there forever.

\section{Domain Growth by First-Passage Resetting}
\label{sec:growth}

We now turn to a different aspect of first-passage resetting---the growth of
a domain as a result of a diffusing particle that reaches the resetting
boundary and causes this boundary to recede by a specified amount at each
resetting event.  Moving boundaries typically arise at the interface between
two thermodynamic phases that undergo a first-order phase
transition~\cite{crank1987free,rubinvsteuin2000stefan}.  In this case, the
interface moves continuously as the stable phase grows into the unstable
phase.  A simple example is water freezing at the interface between water and
air, when the air temperature is held below $0^\circ$C.  A layer of ice grows
on top of the water as heat is transported away from the ice-water interface.
In these types of systems, the temperature field evolves by diffusion and the
movement of the interface is determined by the heat flow at the interface.

In contrast, for a growth process that is induced by first-passage resetting,
a single diffusing particle is discontinuously reset to a distant location
when the boundary is reached.  Concomitantly, the motion of the interface is
intermittent and discontinuous when the interface recedes by a finite distance
upon hitting.  A related behavior also occurs in the absence of resetting:
returning to the situation depicted in ~Fig.~\ref{fig:reln}~(a), a boundary
that is initially at $L$ moves to $2L$ when it is hit, and then to $3L$, etc.
This interface position clearly moves discontinuously and its position moves
as $\sqrt{4Dt/\pi}$.

In the next paragraphs, we study the interface motion and related properties
in the presence of first-passage resetting, when the domain of interest is
either the semi-infinite line or the finite interval.  We find a variety of
growth laws that depend on how far the boundary recedes at each resetting
event.

\subsection{Expanding Semi-Infinite Geometry}

Suppose that the diffusing particle starts at the origin and diffuses in the
range $[-\infty,L_n]$, with $L_n>0$.  Each time the particle reaches $L_n$,
the particle is reset to the origin and the interface moves forward by a
specified amount $\delta L_n$ so that $L_{n+1}=L_{n}+\delta L_{n}$.  Since
the resetting events occur at separated discrete times, it is convenient to
index the position of the interface by $n$, the number of resetting events.
We consider two natural cases: additive and multiplicative interface growth.

\subsubsection{Additive growth: $L_n=L_{n-1}+L$.}

In this case, the right boundary starts at $L$ and then moves to $2L$ at the
first reset event, then to $3L$, etc. We make use of the simple relation
between diffusion with resetting and free diffusion as shown in
Fig.~\ref{fig:reln}. By this equivalence, the probability for $n$ reset
events to occur in the time range $[0,t]$ equals the probability that free
diffusion travels further than $L_n=n(n+1)L/2$ but no further than
$L_{n+1}=(n+1)(n+2)L/2$ in $[0,t]$; that is, the maximum of the freely
diffusing particle is located in the range $[L_n,L_{n+1}]$. So, writing again
$\mathcal{M}(t)$ for the average of the maximum $M(t)$, one has
\begin{align}
  \sum_{n\geq 0}\,L_n\, P\left(N(t)=n\right)\; \leq\; \sum_{n\geq 0}\,\int_{L_n}^{L_{n+1}}\,dm&\,m\,P\left(M(t)=m\right)\; \leq\; \sum_{n\geq 0}\,L_{n+1}\, P\left(N(t)=n\right)\,,\nonumber
 \end{align}
that is 
 \begin{align}
 \label{eq:genM}
   \sum_{n\geq 0}\,L_n\, P\left(N(t)=n\right)\; \leq \;
   &\mathcal{M}(t)\; \leq\; \sum_{n\geq 0}\,L_{n+1}\, P\left(N(t)=n\right)\,.
\end{align}
This leads to
\begin{align}
  \frac{L}{2}\left[\left\langle N^2\right\rangle + \left\langle
  N\right\rangle \right]
  \; \leq \; &\mathcal{M}(t)\; \leq\; \frac{L}{2}\left[\left\langle N^2\right\rangle + 3\left\langle N\right\rangle + 2 \right]\,,
\end{align}
where we write $N$ for $N(t)$ to simplify the notation, from which it follows
that
\begin{align}
  \label{N2}
  \left\langle N^2\right\rangle \simeq 4\sqrt{\frac{Dt}{\pi L^2}} \quad \text{or} \quad \sqrt{\left\langle N^2\right\rangle} \simeq 2\left[\frac{Dt}{\pi L^2}\right]^\frac{1}{4}
\end{align}
Note that we do not obtain directly the average number of reset events
$\mathcal{N}(t)\equiv \left\langle N \right\rangle$ from \eqref{N2}, but only
that it scales as $t^{1/4}$. However, we can derive $\mathcal{N}(t)$ by
exploiting the renewal structure of the problem in the Laplace domain
(see~\ref{ap:sigg}) and find
\begin{align}
    \mathcal{N}(t) \simeq \sqrt{\frac{\pi}{2}}\,\frac{1}{\Gamma(5/4)}
  \times \left(\frac{t}{\tau}\right)^{1/4}\,,
\end{align}
where $\tau=L^2/D$ is the diffusion time.

The $t^{1/4}$ scaling is to be compared with the $t^{1/2}$ scaling when the
boundary is moving through first-passage dynamics but without resetting. When
resetting occurs, the number of encounters with the boundary is reduced
because after each reset, the boundary is further away.    As one might
expect, this boundary recession leads to an anomalously slow
interface growth.

\subsubsection{Multiplicative growth: $L_n=\alpha L_{n-1}$, $\alpha>1$.}

The approach given above can be applied to multiplicative interface
recession.  That is, upon the first resetting, the initial boundary at $x=L$
moves to $x=\alpha L$.  In the next resetting, the boundary moves from
$x=\alpha L$ to $x=\alpha^2 L$, etc.  For this recession rule,
Eq.~\eqref{eq:genM} remains valid, with $L_n=\alpha^n L$. Thus, we have
 \begin{align}
   \sum_{n\geq 0}\,\alpha^n L\, P\left(N(t)=n\right)\; \leq \;
   &\mathcal{M}(t)\; \leq\; \sum_{n\geq 0}\,\alpha^{n+1}L\, P\left(N(t)=n\right)\,,
\end{align}
from which
\begin{align}
  \left\langle \alpha^{N}\right\rangle L\; \leq \;
  &\mathcal{M}(t)\; \leq\; \left\langle \alpha^{N+1}\right\rangle L\,.
\end{align} 
These inequalities suggest that $\mathcal{N}(t)$ scales as
$\ln (t/\tau)\,/\,2 \ln \alpha$.  From the exact Laplace transform approach
(see~\ref{ap:sigg}), we also find the same prefactor in the scaling of
$\mathcal{N}(t)$ with $t$.  Thus we conclude that
\begin{align}
  \mathcal{N}(t)\simeq  \ln (t/\tau)\,/\,2 \ln \alpha\,.
  \label{eq:Nmultres}
\end{align}

After $n$ resets, the boundary is located at $\alpha ^n L$.  We also checked
numerically that the distribution of $N(t)$ is concentrated sufficiently
tightly around its average value $\mathcal{N}(t)$ so that
$\langle \alpha ^N\rangle\sim \alpha^{\langle
  N\rangle}=\alpha^{\mathcal{N}}$.  As a result, the average position of the
boundary at time $t$ scales as
\begin{align}
  \alpha ^{\mathcal{N}(t)}\, L \simeq  \sqrt{t/\tau}\,L\,.
  \label{eq:PosBar}
\end{align}
Thus the boundary moves as $t^{1/2}$, which is faster than in the additive
case, where the boundary moves as $t^{1/4}$.  Despite a smaller number of
reset events, each of these moves the boundary far enough for the overall
motion to be almost as fast as in the case of first-passage growth without
resetting, with the difference being only a factor of 2.

\subsection{Expanding Interval}

We now study the case where a diffusing particle is confined to a finite and
growing interval $[0,L_n]$, with a reflecting boundary condition at $x=0$.
Each time the particle reaches the right boundary at $x=L_n$, the particle is
instantaneously reset to $x=0$, while the position of the boundary recedes by
a specified amount.  We want to understand how the interval grows with time
and related statistical properties of this process.  We first give the formal
result for an arbitrary dependence of $L_n$ on $n$ and then specialize to the
additive case where $L_n=nL$.

We again start with the analog of Eq.~\eqref{eq:Rns} for the finite domain,
namely, the Laplace transform of the probability to reset for the
$n^{\text th}$ time at $t$:
\begin{align}
  \label{eq:Rns-EI}
  \widetilde{R}_{n}(s) &= \widetilde{R}_{n-1}(s) \sech\big(\sqrt{s/D}\,L_n\big)
                         =\prod_{m=1}^n \sech\big(\sqrt{s/D}\,L_m\big) \,.
\end{align}
Here $\sech\big(\sqrt{s/D}\,L_n\big)$ is the Laplace transform of the
first-passage probability to the right boundary of the finite interval
$[0,L_n]$~\cite{redner2001guide}.  Similarly, the average number of resetting
events obeys the renewal equation
\begin{subequations}
\begin{align}
  \label{eq:Ntint}
\mathcal{N}(t)= 0\times Q(L,t) +  \sum_{n=1}^\infty n\int_0^t dt'\, Q(L_{n+1},t-t')R_{n}(t') \,,
\end{align}
where $Q(L,t)$ is now the survival probability of a diffusing particle in the
finite interval $[0,L_n]$, with reflection at $x=0$ and absorption at
$x=L_n$. In the Laplace domain Eq.~\eqref{eq:Ntint} becomes
\begin{align}
  \label{eq:Nsxxx}
  \widetilde{\mathcal{N}}(s)= \sum_{n=0}^\infty n
  \widetilde{Q}(L_{n+1},s)\widetilde{R}_{n}(s) \,.
\end{align}
\end{subequations}
Substituting in $\widetilde{Q}(L_n,s)=[1-\widetilde{F}(L_n,s)]/s$ and
Eq.~\eqref{eq:Rns} into the above equation gives
\begin{align}
  \widetilde{\mathcal{N}}(s)=& \sum_{n=0}^\infty \frac{n}{s}\,
     \left[1-\sech\big(\sqrt{s/D} L_{n+1}\big)\right] \prod_{m=1}^n \sech\big(\sqrt{s/D}L_m\big) \nonumber\\
=& \sum_{n=0}^\infty \frac{n}{s}\left[ \widetilde{R}_n(s)-\widetilde{R}_{n+1}(s)\right]
=\frac{1}{s}\sum_{n=1}^\infty \widetilde{R}_n(s)\nonumber\\
=&\frac{1}{s}\sum_{n=1}^\infty \prod_{m=1}^n \sech\big(\sqrt{s/D}L_m\big)\label{eq:Nsum}
\end{align}

To extract the asymptotic behavior of $\mathcal{N}(t)$, we now focus on the
additive case where $L_n=nL$; that is, the boundary recedes by a fixed
distance $L$ after each resetting event.  Using the dimensionless coordinate
$\ell=\sqrt{sL^2/D}$, Eq.~\eqref{eq:Nsum} now gives
\begin{align}
  \widetilde{\mathcal{N}}(s)
  &= \frac{1}{s}\sum_{n=1}^\infty \prod_{m=1}^n \sech\big(\sqrt{s/D}\,\, m L\big)
 = \frac{1}{s}\sum_{n=1}^\infty \prod_{m=1}^n \sech(m\,\ell)\nonumber\\
 &\approx \frac{1}{s}\sum_{n=1}^\infty \exp\left\{\int_0^n dm \ln\big[\sech(m \,\ell)\big]\right\}\nonumber \\
  &\approx \frac{1}{s}\sum_{n=1}^\infty \exp\left\{- \frac{1}{\ell}
    \left[-\frac{1}{2}\, \ell^2 n^2-\frac{1}{2} \text{Li}_2\left(-e^{2 \ell n}\right)
    -\ell n \ln 2-\frac{\pi ^2}{24} \right] \right\}\nonumber \\
  &\approx \frac{1}{s \ell}\int_{0}^\infty du\,
    \exp\left\{- \frac{1}{\ell}\left[-\frac{1}{2} u^2-\frac{1}{2}
    \text{Li}_2\left(-e^{2 u}\right)-u \ln 2-\frac{\pi ^2}{24} \right] \right\}\,.\nonumber
\end{align}
where $\text{Li}_2(x)$ is the polylogarithm function of order $2$. This integral can now be computed in the small $s$ limit using the
saddle-point approximation (see \ref{ap:saddlepoint}) and the final result is:
\begin{align}
  \widetilde{\mathcal{N}}(s)  &\approx \frac{1}{s \ell}\int_0^\infty du\, \exp\left(-\frac{ u^3}{6}\right)\nonumber\\
    &\approx \Gamma \left(\frac{4}{3}\right)\,\frac{1}{s}\,   \left(\frac{6}{\ell^2}\right)^{1/3}+ o\left(\frac{1}{s^{4/3}}\right)\,.
    \label{eq:saddlepoint}
\end{align}
The Laplace inversion of the above expression gives the average number of
resetting events up to time $t$ in the $t\to\infty$ limit:
\begin{align}
  \mathcal{N}(t) \simeq \left(\frac{6t}{\tau}\right)^{1/3}+o\left(t^{1/3}\right)\,.
  \label{eq:Navgint}
\end{align}
This result implies that the length of the interval also grows as $t^{1/3}$.
For determining the standard deviation (see below), we also need the next
correction to the asymptotic behavior.  Numerically, we find that
$\mathcal{N}\simeq (6t/\tau)^{1/3}+C_1$ where $C_1= -0.8$.

The $t^{1/3}$ dependence of $\mathcal{N}(t)$ can be understood in a simple
way.  The mean time for a particle, which starts at $x=0$, to reach the
boundary at $x=L_n$ is
$L_n^2/2D= (nL)^2/2D\equiv n^2\tau/2$~\cite{redner2001guide}.  If the
particle is immediately reset to the origin each time the boundary is
reached, then the time required for $\mathcal{N}$ reset events is
$\sum^{\mathcal{N}} n^2 \tau/2 \simeq \mathcal{N}^3\tau/6$.  This gives
$\mathcal{N}\simeq (6t/\tau)^{1/3}$.

The second moment of the probability distribution for the number of
encounters is:
\begin{align}
  \langle \widetilde{{N}}^2(s)\rangle
  = \sum_{n=0}^\infty \frac{n^2}{s}\left[ \widetilde{R}_n(s)-\widetilde{R}_{n+1}(s)\right] = \sum_{n=1}^\infty \frac{2n-1}{s} \widetilde{R}_n(s)
\end{align}
By following similar steps as those to compute $\mathcal{N}$, we obtain the
following result for fixed $n$ and small $s$:
\begin{align}
    \langle \widetilde{{N}}^2(s)\rangle
  &\approx  \frac{1}{s\ell}\int_{0}^\infty  du\,\left(\frac{2u}{\ell}-1\right)
  \exp\left\{- \frac{1}{\ell}\left[-\frac{1}{2} u^2-\frac{1}{2}
   \text{Li}_2\left(-e^{2 u}\right)-u \ln 2-\frac{\pi ^2}{24} \right] \right\}\,.\nonumber 
\end{align}
This integral can now be computed in the small-$s$ limit using the saddle-point approximation:
    \begin{align}
  \langle \widetilde{{N}}^2(s)\rangle  &\approx \frac{1}{s \ell}\int_0^\infty du\, \left(\frac{2u}{\ell}-1\right)\,\exp\left(-\frac{\ell\, u^3}{6}\right)\nonumber\\
    &\approx \Gamma \left(\frac{5}{3}\right)\frac{1}{s}\left(\frac{6}{ \ell^2}\right)^{2/3}+o\left( \frac{1}{s^{5/3}}\right)\,.
\end{align}
Performing a Laplace inversion of the above expression gives:
\begin{align}
    \langle \mathcal{N}^2(t)\rangle  \simeq \left(\frac{6\,t}{\tau}\right)^{2/3} + o\left(t^{2/3}\right)\, .
\end{align}
Numerically, we find that the next correction is $C_2\, (6\, t/\tau)^{1/3}$
with $C_2\approx -1.47$. Hence, the standard deviation grows as
$\sqrt{\langle {N}^2(t)\rangle-\langle {N}(t)\rangle^2}\approx \sqrt{C_2-2\,
  C_1}\,(6\, t/\tau)^{1/6}$.

\begin{figure}[ht]
     \centerline{\includegraphics[width=0.475\textwidth]{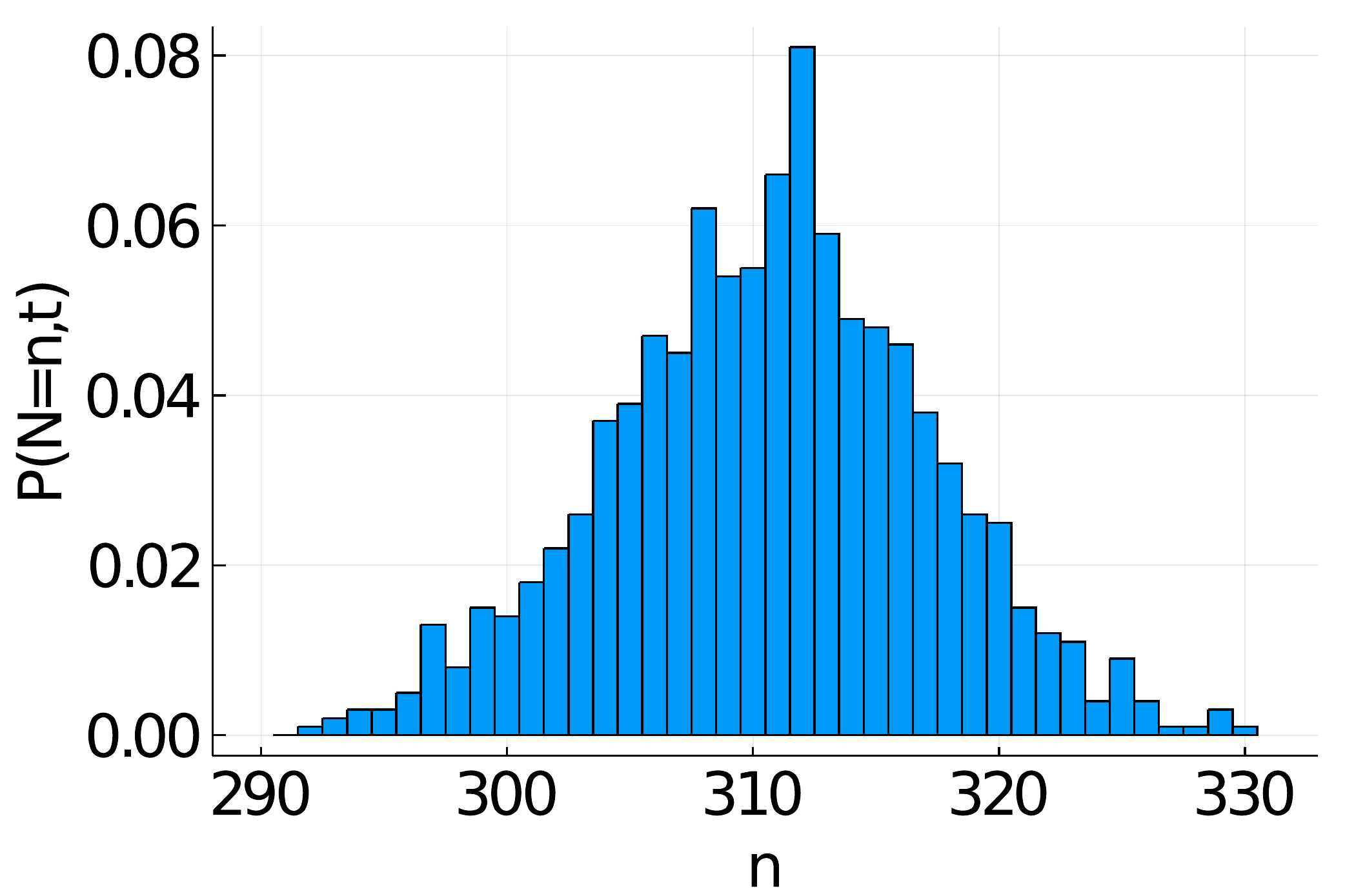}}
     \caption{Numerical simulation results for $P(N\!=\!n,t)$ for the
       expanding interval.  The initial interval length $L=1$ and the length
       grows by 1 after each resetting event.  The distribution is shown at
       $t=10^9$ for $1000$ walkers.  The diffusion constant was set to
       $\Delta x^2/(2\Delta t)=5\times10^{-3}$. }
      \label{fig:Ndist}
\end{figure}

Numerical simulations of this growth process (Fig.~\ref{fig:Ndist}) show that
the distribution of $\mathcal{N}(t)$ is highly localized around its average
value and decreases rapidly as one moves away from the maximum.  While we do
not know how to compute the full distribution analytically, we can determine
the tails of the distribution by a simple extremal
argument~\cite{fisher1966shape,krapivsky2010kinetic}.  For notational
simplicity we take $L=1$ and $D=1$.  From the time dependence of the average
value of $\mathcal{N}(t)$ (Eq.~\eqref{eq:Navgint}), we posit that the natural
scaling variable is $z\equiv n/ \mathcal{N}(t)\simeq n/t^{1/3}$.  We further
assume that the distribution can be expressed in the scaling form
$P({N}=n,t)\propto f(z)$ that decays as a stretched exponential for both
$z\to\infty$ and $z\to 0$.  That is, $f(z)=\exp(-z^a)$, where $a>0$, for
$z\to\infty$ and $f(z)=\exp(-z^{b})$, where $b<0$, for $z\to 0$.

Consider now the extreme event in which the particle always moves towards the
resetting boundary up to time $t$.  This event occurs with probability
$2^{-t} \sim e^{-t}$.  For this directed motion, the particle
requires 1 time step to first reach the boundary, 2 time steps to reach it
for a second time, 3 time steps for a third time, etc.  This leads to the
total number $n$ of encounters with the boundary that is determined by
$\sum_{k=1}^n k=t$.  Hence, $n\simeq \sqrt{2t}$.  In terms of our scaling
function, the probability to reach the boundary $\sqrt{2t}$ times occurs with
probability $e^{-t^{a/6}}$.  Equating this with $e^{-t}$ gives $a=6$.

For the small-$z$ tail, we focus on the situation where the boundary is
encountered as little as possible.  This extremal event is achieved by a
random walk that alternately and deterministically moves one step left, then
one step right, etc.  In the case the boundary is encountered once and only
once.  This event again occurs with probability $2^{-t} \simeq e^{-t}$.  On the
other hand, this event of a single boundary encounter corresponds to the
scaling variable $z= 1/t^{1/3}\to 0$, and thus occurs with probability
$e^{-t^{-b/3}}$.  Equating these two asymptotic forms of the distribution
gives $b=-3$. In summary, we find the following asymptotics:
\begin{align}
    P(N=n,t) \simeq\begin{cases}
      e^{-(n/\mathcal{N}(t))^6} \qquad\ n \rightarrow \infty\,, \\
      e^{-(n/\mathcal{N}(t))^{-3}}\qquad n \rightarrow 0\,.
    \end{cases}
\end{align}
Because the exponent values in the scaling forms are fairly large, it does
not seem possible to verify these asymptotic behaviors numerically.

\section{Summary and Discussion}

We presented the concept of first-passage resetting, in which a random walk
is reset to its starting point whenever it reaches a specified location.
This situation contrasts with constant-rate resetting in which a random walk
is reset to its starting point at a fixed rate.  In the simple case of a
semi-infinite line, $[-\infty,L]$ with $L>0$, the particle diffuses freely
and is reset to the origin whenever it reaches $L$.  The resulting
probability distribution has dramatically different behavior depending on
whether $0<x<L$ or $x<0$.  In the former case, the distribution has a simple
linear profile that arises from the balance between flux leaving at the reset
point and the flux being reinjected at $x=0$.  In the latter case, the
probability distribution reduces to free diffusion in the presence of a
reflecting boundary.  We derived this result analytically and also via a path
decomposition that is reminiscent of the image method.

In the finite interval geometry, we defined an optimization problem that
describes, in a schematic way, aspects of the repeated breakdown of a driven
mechanical system.  The operation domain of the system is a finite interval;
this interval could be interpreted as the RPM range of an engine.  The
resetting boundary corresponds the system reaching its operating limit or
maximum RPMs, after which a breakdown occurs and the system has to be
restarted from scratch.  The control parameter is the bias velocity (not to
be confused with the RPM of the engine), which may either drive the system
towards breakdown or towards minimal-level operation.  We showed that there
exists an optimal bias velocity that optimizes the performance of the system.
This optimum balances the gain by operating close to $x=L$ while minimizing
the number of breakdowns.  A similar physical picture arises if breakdown is
accompanied by a random delay before restarting the system or by extending to
a two-dimensional geometry.

We also studied a variety domain growth phenomena that are driven by
first-passage dynamics with resetting. When each resetting event moves the
boundary by a fixed amount, the boundary recedes as $t^{1/4}$ and as
$t^{1/3}$ for the semi-infinite geometry and the finite interval,
respectively. In the semi infinite geometry, if the boundary position grows
by a factor $\alpha > 1$ with each resetting event, then the interface moves
much more quickly, as $t^{1/2}$.  The case where the boundary moves by a
fixed amount at each resetting is actually a version of the internal
diffusion-limited aggregation problem for which there is extensive literature
that has focused on the geometrical properties of the growing domain (see,
e.g.,
\cite{meakin1986formation,lawler1992internal,moore2000internal,jerison2012logarithmic}).
We instead focused on the rich dynamical aspects of the model and we
suggest, based on the correspondence with internal diffusion-limited
aggregation, that it will be worthwhile to treat our first--passage resetting
in a finite two-dimensional domain.

There are a variety of extensions of the optimization problem may be worth
exploring.  First, the control strategy could be finer than simply a uniform
bias velocity~\cite{lunz2020}.  More realistically, one could also associate a cost to a
strategy that becomes more expensive as the control mechanism becomes more
sophisticated.  For instance, it would be natural to turn on a bias velocity
\emph{away} from the breakdown point when the system is very close to
breakdown.  It would also be useful to identify the optimal region over which
the particle experiences a bias (both toward and away from the breakdown
point).  In addition to a more refined control strategy, other simple
geometries may be worthwhile to study.  One such example is a one-dimensional
interval with a first-passage resetting mechanism at each end of the interval
and with a different cost in reaching each end.  First-passage resetting in a
bounded planar geometry with a cost function that depends on the hitting
angle of the boundary might be another geometry that will be worthwhile to
study.
 
Given the rich behavior exhibited by first-passage resetting, it should also
be worthwhile to investigate both extensions of the basic model and
applications.  An example of the former is the Fleming-Viot branching
process, in which there are $N+1$ particles and when one of them resets, it
resets to one of the positions of the remaining $N$
particles~\cite{fleming1979some,burdzy1996configurational,burdzy2000fleming,grigorescu2004hydrodynamic,grigorescu2006tagged}.
More generally, the first-passage resetting in the presence of multiple
diffusing particles could lead to new phenomenology.  On another note,
applications also exist in cash flow management: cash levels in a large firm
are sometimes modeled as a diffusion process in which one wishes to have cash
fully invested in profitable ventures, while at the same time keeping enough
cash available so as to avoid being indebted
~\cite{harrison1983impulse,buckholtz1983analysis}.  These types of problems
seem to be ripe for further exploration.

\section{Acknowledgments}
BBs research at the Perimeter Institute was supported in part by the
Government of Canada through the Department of Innovation, Science and
Economic Development Canada and by the Province of Ontario through the
Ministry of Colleges and Universities. JRFs research at Columbia University
was supported by the Alliance Program.  SR thanks Paul Hines for helpful
conversations and financial support from NSF grant DMR-1910736.  We thank one
of the referees for providing helpful suggestions about possible extensions
of the basic optimization problem.

\appendix
\section{Laplace transform approach in the semi-infinite geometry}
\label{app-a}

\subsection{Spatial probability distributions}

The probability distribution $P(x,t)$ of the diffusing particle at time $t$
on the semi-infinite line $x\leq L$, can be obtained in several ways. We
presented in the main text a path transformation approach and we detail here
the Laplace transform approach (see also~\cite{bruyne2020firstpassage}). We
first partition the trajectory according to the number of reset events up to
time $t$.  Between consecutive resets, the particle undergoes free diffusion
with an absorbing boundary at $x\!=\!L$.  This part of the motion is described by
the free propagator
\begin{align}
  \label{G}
  G(x,L,t) =\big[e^{-x^2/4Dt} - e^{-(x-2L)^2/4Dt}\big]/\sqrt{4\pi Dt}\,,
\end{align}
which can be computed, for example, by the image
method~\cite{feller2008introduction,redner2001guide}.  Summing over all
numbers of reset events, the spatial probability is determined by
\begin{subequations}
  \label{eq:renEq}
\begin{align}
  \label{Pxt}
    P(x,t) = G(x,L,t)+   \sum_{n\geq 1}\, \int_0^{t} dt '\, F_n(L,t')\,  G(x,L,t\!-\!t')\,.
\end{align}
Equation~\eqref{Pxt} states that for the particle to be at $x$ at time $t$,
it either: (i) must never hit $L$, in which case its probability distribution
is just $G(x,L,t)$, or (ii), the particle first hits~$L$ for the $n^{\rm th}$
time at $t'<t$, after which the particle restarts at the origin and then
propagates to $x$ in the remaining time $t-t'$ without hitting~$L$ again.
The equivalent way of writing Eq.~\eqref{Pxt} in a renewal fashion is:
\begin{align}
\label{eq:Pxtb}
P(x,t) =  G(x,L,t) + \int_0^t dt' \,F_{1}(L,t')P(x,t-t')\,.
\end{align}
\end{subequations}
The first term accounts for the particle never reaching $x=L$, while the
second term accounts for the particle reaching $x=L$ at time $t'$, after
which the process starts anew from $x(t')=0$ for the remaining time
$t-t'$. Note that this is a renewal equation in the sense that the second
term contains the full propagator $P(x,t-t')$ and not the free propagator
$G(x,t-t')$, thereby accounting for any number of resetting events in the
time interval $[t',t]$.

To solve for $P(x,t)$ we again treat the problem in the Laplace domain.
While we can find the solution from the Laplace transform of Eq.~\eqref{Pxt},
the solution is simpler and more direct from the Laplace transform
of~\eqref{eq:Pxtb}:
\begin{subequations}
\begin{align}
   \widetilde{P}(y,s) &=  \widetilde{G}(y,\ell,s)
                        + \widetilde{F}_1(\ell,s)
                        \widetilde{P}(y,s),
\end{align}
with
\begin{align*}
\widetilde{G}(y,\ell,s)=\left[
  e^{-|y|} -  e^{-|y-2\ell|}\right]/\sqrt{4Ds}\,,
\end{align*}
the Laplace transform of $G(x,L,t)$, where we have introduced the scaled
coordinates $y\equiv x\sqrt{s/D}$ and $\ell\equiv L\sqrt{s/D}$.  Solving for
$ \widetilde{P}(y,s)$ yields:
\begin{align}
 \label{Pxs-sol}
  \widetilde{P}(y,s) &=   \frac{\widetilde{G}(y,\ell,s)}
        {1-\widetilde{F}_1(\ell,s)}
        =  \frac{1}{\sqrt{4Ds}}\,\frac{\left[
  e^{-|y|} -  e^{-|y-2\ell|}\right]}
 {1-e^{-\ell}}\,.
\end{align}
\end{subequations}
To invert this Laplace transform, we separately consider the cases
$0\leq y\leq \ell$ and $y<0$.  In the former, we expand the denominator in a
Taylor series to give
\begin{subequations}
\label{P<}
  \begin{align}
 \label{Pxs-sol<}
  \widetilde{P}(y,s)
    & = \frac{1}{\sqrt{4Ds}}\,
  \left[ e^{-y}-e^{-(2\ell-y)}\right]\,
      \sum_{n\geq 0} e^{-n\ell}\nonumber\\
    &=  \frac{1}{\sqrt{4Ds}}\,\sum_{n\geq 0}
      \left[ e^{-(y+n\ell)}
      -e^{-[(n+2)\ell-y]}\right]\,, \qquad 0<y\leq \ell\,,
\end{align}
from which 
\begin{align}
 \label{Pxt-sol<}
  P(x,t)
  &=  \frac{1}{\sqrt{4\pi Dt}}\,\sum_{n\geq 0}\left[ e^{-(x+nL)^2/4Dt}
   - e^{-[x-(n+2)L]^2/4Dt}\right],\qquad 0<x\leq L\,.
\end{align}
\end{subequations}
In the case of $y<0$, $\widetilde{P}(y,s)$ in Eq.~\eqref{Pxs-sol} is factorizable:
\begin{subequations}
\begin{align}
  \widetilde{P}(y,s) &= \frac{1}{\sqrt{4Ds}}\left[\frac{e^{y}- e^{y-2\ell}}{1-e^{-\ell}}\right]
  =  \frac{1}{\sqrt{4Ds}}\left[e^{y}+ e^{(y-\ell)}\right],
\end{align}
and this latter form can be readily inverted to give:
\begin{align}
  P(x,t) = \frac{1}{\sqrt{4\pi Dt}}\,
  \left[e^{-x^2/4Dt}+ e^{-(x-L)^2/4Dt}\right]\quad x<0\,.
\end{align}
\end{subequations}

\subsection{Average number of resets}

The average number of resets that occur up to time $t$ may also be derived by
using the Laplace transform approach.  The probability for $n$ resets to
occur by time $t$ equals the probability to have at least $n$ resets minus
the probability to have at least $n+1$ resets:
\begin{subequations}
\begin{align}
\begin{split}
P\left(N(t)\!=\!n\right) &= P\left(N(t)\!\geq\! n\right)-P\left(N(t)\!\geq\! n+1\right)\, ,\\
&= \int_0^{t}\! dt '\, F_n(L,t')\,-\int_0^{t}\! dt '\, F_{n+1}(L,t')\,.
\end{split}
\end{align}
Using our earlier result that $F_n(L,t)=F_1(nL,t)$, we have
\begin{equation}
P\left(N(t)\!=\!n\right)=\mathrm{erf}\left(\frac{(n+1)L}{\sqrt{4Dt}}\right)-\mathrm{erf}\left(\frac{nL}{\sqrt{4Dt}}\right)\,,
\end{equation}
\end{subequations}
where $\mathrm{erf}$ is the Gauss error function.

We can compute the average number of reset events,
$\mathcal{N}(t)\equiv \left\langle N(t) \right\rangle$,
from~\eqref{eq:ndist}, but it is quicker to use a renewal equation approach.
Here we can write
\begin{subequations}
\begin{align}
  \label{eq:navgb}
  \mathcal{N}(t) = \int_0^t dt' \,F_{1}(L,t-t')\big[1+\mathcal{N}(t')\big]\,.
\end{align}
Equation~\eqref{eq:navgb} states that to have $\mathcal{N}$ reset events up
to time $t$, $\mathcal{N}-1$ reset events must have occurred at some earlier
time $t'<t$ and then one more reset event occurs exactly at time $t$. Taking
the Laplace transform of~\eqref{eq:navgb} gives
\begin{align}
\label{eq:Nsi}
\widetilde{\mathcal{N}}(s) &= \frac{\widetilde{F}_{1}(L,s)}{s\,(1-\widetilde{F}_{1}(L,s))} 
   = \frac{e^{-\ell}}{s(1-e^{-\ell})}\,,
\end{align}
\end{subequations}
where again $\ell = L\sqrt{s/D}$.  We extract the long-time behavior of the
average number of reset events by taking the $s\rightarrow 0$ limit and then
Laplace inverting this limiting expression.  We thus find
\begin{align}
  \mathcal{N}(t) \simeq \sqrt{4Dt/\pi L^2}\,.
\end{align}

\section{ Solution to the convection-diffusion equation}
\label{app:applybc}

The general solution to Eq.~\eqref{eq:difffree} is
\begin{align}
\label{eq:generalsol}
  \widetilde{c_0}(x,s) = e^{\P}(A e^{wx}+B e^{-wx})\,,
\end{align}
where $\P=vx/2D$, $w = \sqrt{v^2+4Ds}/2D$, and $A, B$ are integration
constants.  To determine $A$ and $B$, we apply the boundary conditions that
accompany Eq.~\eqref{eq:generalsol} to give the linear system
\begin{align}
     \begin{cases}
     v (A+B)- \frac{1}{2}v(A+B) -AD  w + BD w= 1\,, \\
     e^{vL/(2 D)} \left(A e^{L w}+B e^{-L w}\right) = 0\,,
     \end{cases}
\end{align}
whose solution is
\begin{align}
     \begin{cases}
     A=-\dfrac{2}{2 D w e^{2 L w}+2 D w+v e^{2 L w}-v}=-\dfrac{e^{-L w}}{2 D w \cosh(Lw)+v\sinh(Lw)}\,,\\[0.7em]
     B=\dfrac{2 e^{2 L w}}{2 D w e^{2 L w}+2 D w+v e^{2 L w}-v}=\dfrac{e^{L w}}{2 D w \cosh(Lw)+v\sinh(Lw)}\,.
     \end{cases}
\end{align}
We define $\mathcal{W}\equiv 2 D w \cosh(Lw)+v\sinh(Lw)$, from which
$A=-e^{-Lw}/\mathcal{W}$ and $B=e^{Lw}/\mathcal{W}$. Substituting these
constants back into the general solution Eq.~\eqref{eq:generalsol} leads to
Eq.~\eqref{eq:c0}.

\section{First-passage resetting in the annular geometry}
\label{app:fpr2d}

For the convection-diffusion equation in two dimensions with a radial drift
$v/r$; that is, Eq.~\eqref{eq:c2da} without the delta-function term, the
general solution in the Laplace domain is~\cite{abramowitz1988handbook}
\begin{align}
\label{eq:cABfp}  
  \widetilde{c}_0(x,s) &= x^{\Pe/2}\big[A\,\I_{\Pe/2}(\sqrt{s}x)
                         +B\, \K_{\Pe/2}(\sqrt{s}x)\big]\,,
\end{align}
where $A$, and $B$ are integration constants, and $I_{\nu}(x)$ and
$K_{\nu}(x)$ are the modified Bessel functions of the first and second kind,
respectively. The subscript $0$ refers to the
concentration without flux re-injection. Imposing the boundary conditions
that accompany Eq.~\eqref{eq:c2da} leads to a linear system to solve for $A$
and $B$:
\begin{align}
\begin{cases}
 A I_{{\Pe}/{2}}\left(\sqrt{s}\right)+B
   K_{{\Pe}/{2}}\left(\sqrt{s}\right) &= 0\,, \\
   \sqrt{s} x_0^{1+{\Pe}/{2}} \left(B K_{1+\Pe/2}\left(\sqrt{s}
   x_0\right)-A I_{1+\Pe/2}\left(\sqrt{s} x\right)\right)&=1/(2\pi)\,.
\end{cases}
\end{align}
whose solution is:
\begin{align}
         \begin{cases}
     A=-\dfrac{x_0^{-(1+\Pe/2)}
   K_{{\Pe}/{2}}\left(\sqrt{s}\right)}{2\pi\sqrt{s} \left( 
   K_{{\Pe}/{2}}\left(\sqrt{s}\right) I_{1+\Pe/2}\left(\sqrt{s}
   x_0\right)+  I_{{\Pe}/{2}}\left(\sqrt{s}\right)
   K_{1+\Pe/2}\left(\sqrt{s} x_0\right)\right)}\,,\\[0.7em]
     B=\dfrac{x_0^{-(1+\Pe/2)}
   I_{{\Pe}/{2}}\left(\sqrt{s}\right)}{2\pi\sqrt{s} \left( 
   K_{{\Pe}/{2}}\left(\sqrt{s}\right) I_{1+\Pe/2}\left(\sqrt{s}
   x_0\right)+  I_{{\Pe}/{2}}\left(\sqrt{s}\right)
   K_{1+\Pe/2}\left(\sqrt{s} x_0\right)\right)}\,.
     \end{cases}
\end{align}
We now define
$\mathcal{W}\equiv x_0^{1+\Pe/2} \sqrt{s}\left[
  K_{{\Pe}/{2}}\left(\sqrt{s}\right) I_{1+\Pe/2}\left(\sqrt{s} x_0\right) +
  I_{{\Pe}/{2}}\left(\sqrt{s}\right) K_{1+\Pe/2}\left(\sqrt{s}
    x_0\right)\right]$.  In terms of this function, we have
$A=K_{{\Pe}/{2}}\left(\sqrt{s}\right)/(2\pi \mathcal{W})$ and
$B= I_{{\Pe}/{2}}\left(\sqrt{s}\right)/(2\pi\mathcal{W})$.  Substituting
these constants back into Eq.~\eqref{eq:cABfp} yields:
\begin{align}
\label{eq:c2dasol}  
\widetilde{c}_0(x,s) = \frac{x^{\Pe/2}\left[I_{{\Pe}/{2}}\left(\sqrt{s}\right)
 K_{{\Pe}/{2}}\left(\sqrt{s} x\right)-K_{{\Pe}/{2}}\left(\sqrt{s}\right)
 I_{{\Pe}/{2}}\left(\sqrt{s} x\right)\right]}{2 \pi \mathcal{W}}\,,
\end{align}
The first-passage probability is obtained by computing the outlet flux at
$x=1$ from the concentration in Eq.~\eqref{eq:c2dasol}:
\begin{align}
\label{eq:F12d}
\widetilde{F}_1(s) &= 2\pi \left( -x\partial_x \widetilde{c}_0(x,s) +
                          \Pe\, \widetilde{c}_0(x,s)
                          \right)\rvert_{x=1}\nonumber \\
     &=2\pi \left( -\partial_x \widetilde{c}_0(x,s)
       \right)\rvert_{x=1}\nonumber \\
     &=\frac{\sqrt{s} I_{{\Pe}/{2}}\left(\sqrt{s}\right)
   K_{1-{\Pe}/{2}}\left(\sqrt{s}\right)+\sqrt{s}
   I_{1-{\Pe}/{2}}\left(\sqrt{s}\right)
   K_{{\Pe}/{2}}\left(\sqrt{s}\right)}{\mathcal{W}}=\frac{1}{\mathcal{W}}\,, 
\end{align}
where we used the absorbing boundary condition to go to the second line.

On the other hand, the survival probability ${Q}(t)$ is defined as the
integral of $2\pi x\, \widetilde{c}_0(x,t)$ over the interval
$[x_0,1]$. Alternatively, it can be computed as the probability of not having
hit the absorbing boundary until time $t$: $Q(t)=1-\int_0^t dt' F_1(t')$,
which in the Laplace domain translates to
$\widetilde{Q}(s) = (1 -\widetilde{F}_1(s))/s$.  Using Eq.~\eqref{eq:F12d},
we obtain:
\begin{align}
\label{eq:Q2d}
\widetilde{Q}(s) = \frac{1}{s}\left(1- \frac{1}{\mathcal{W}}\right)\,.
\end{align}
When there is re-injection of the outgoing flux, the concentration obeys the
renewal equation below. In the Laplace domain and using the form for
$\widetilde{c}_0$ obtained above, we find:
\begin{align}
\label{eq:csp2d}
\widetilde{c}(x,s)= \frac{\widetilde{c}_0(x,s)}{1-\widetilde{F}_1(s)}=\frac{x^{\Pe/2} \left[\I_{{\Pe}/{2}}\left(\sqrt{s}\right)
\K_{{\Pe}/{2}}\left(\sqrt{s} x\right)-\K_{{\Pe}/{2}}\left(\sqrt{s}\right)
 \I_{{\Pe}/{2}}\left(\sqrt{s} x\right)\right]}{2\pi\left(\,  \mathcal{W}-1\right)}\,.
\end{align}
In the $s\to 0$ limit, we find that:
\begin{align}
\begin{split}
&x^{\Pe/2}[\I_{{\Pe}/{2}}\left(\sqrt{s}\right)
\K_{{\Pe}/{2}}\left(\sqrt{s} x\right)-\K_{{\Pe}/{2}}\left(\sqrt{s}\right)
 \I_{{\Pe}/{2}}\left(\sqrt{s} x\right)] \simeq \frac{ 1 - x^\Pe}{\Pe}\,,\\
&\mathcal{W}-1\simeq  \frac{\Pe - \Pe\, x_0^2 + 2 x_0^2\, (-1 + x_0^\Pe) }{2 \Pe\, (2 + \Pe)}\, s\,.
\end{split}
\end{align}

Substituting these asymptotic expressions in Eq.~\eqref{eq:csp2d}, the
coefficient of the term proportional to $1/s$ in $\widetilde{c}(x,s)$ gives
the steady-state probability density in the time domain that is quoted in
Eq.~\eqref{eq:solt2d}.

\section{Expanding semi-infinite geometry: Laplace transforms}
\label{ap:sigg}

We derive here with Laplace transforms the properties for domain growth in
the semi-infinite geometry that was obtained in the main text using a path
transformation.  Suppose that the diffusing particle starts at the origin and
diffuses in the range $[-\infty,L_n]$, with $L_n>0$.  Each time the particle
reaches $L_n$, the particle is reset to the origin and the interface moves
forward by a specified amount $\delta L_n$ so that
$L_{n+1}=L_{n}+\delta L_{n}$.  Since the resetting events occur at separated
discrete times, it is convenient to index the position of the interface by
$n$, the number of resetting events.  We consider two natural cases: additive
and multiplicative interface growth.

\subsection{Additive growth: $L_n=L_{n-1}+L$.}

In this case, the right boundary starts at $L$ and then moves to $2L$ at the
first reset event, then to $3L$, etc.  The probability that the $n^{\text th}$
reset occurs at time $t$, $R_n(t)$, is given by the renewal equation
\begin{subequations}
\begin{align}
\label{eq:Rnt}
R_{n}(t) &= \int_0^t dt'\, R_{n-1}(t-t')\, F(L_n\!=\!n L, t') \,,
\end{align}
where $F(L_1,t)$ is the standard first-passage probability to reach $L_1$
when the particle starts from the origin.  In the Laplace domain
Eq.~\eqref{eq:Rnt} becomes
\begin{align}
\label{eq:Rns}  
\widetilde{R}_{n}(s) &= \widetilde{R}_{n-1}(s)\,  e^{-n\ell} = e^{-n(n+1)\ell/2} \,,
\end{align}
\end{subequations}
with $\ell\equiv\sqrt{sL^2/D}$.

By similar considerations, the average number of resetting events is
\begin{subequations}
\begin{align}
\label{eq:Nt}
\mathcal{N}(t)= 0\times Q(L,T) +  \sum_{n=1}^\infty n\int_0^t dt'\,
Q\big((n\!+\!1)L,t-t'\big)\,R_{n}(t') \,,
\end{align}
which, in the Laplace domain, becomes
\begin{align}
  \label{eq:Ns}
  \widetilde{\mathcal{N}}(s)
  = \sum_{n=0}^\infty n\, \widetilde{Q}\big((n+1)L,s\big)\, \widetilde{R}_{n}(s) \,.
\end{align}
\end{subequations}
Here $Q(L,t)=1-\int_0^T dt F(L,t)$ is the survival probability for a
diffusing particle that starts at the origin to not reach an absorbing
boundary at $L$ within time $T$.

In the Laplace domain, this relation
becomes $\widetilde{Q}(nL,s)=\big[1-\widetilde{F}(nL,s)\big]/s$.  We now
substitute this expression for $\widetilde{Q}$ and the above expression for
$\widetilde{R}_n(s)$ into \eqref{eq:Ns}, and then convert the sum to an integral
to give
\begin{subequations}
\begin{align}
  \widetilde{\mathcal{N}}(s)\approx&
 \int_0^\infty dn\,\, \frac{n}{s} \, \left[1-e^{-(n+1)\ell}\right]\,\, e^{-n(n+1)\ell/2}\nonumber\\
=&\frac{e^{-\ell}}{4\,  s\, \ell} \Big\{\sqrt{2 \pi \ell}\,\, e^{9\,\ell/8}
 \big[2+\text{erf}(\sqrt{\ell/8})-3\, \text{erf}(3 \sqrt{\ell/8})\big]
   +4\, e^{\ell} -4\Big\}
  \nonumber\\[2mm]
   \to&\, \frac{1}{s}\sqrt{\frac{\pi }{2\,\ell}}\qquad s\rightarrow 0\,.
\end{align}
Laplace inverting this expression, the average number of reset events
asymptotically scales as
\begin{align}
  \label{N-14}
    \mathcal{N}(t) \simeq \sqrt{\frac{\pi}{2}}\,\frac{1}{\Gamma(5/4)}
  \times \left(\frac{t}{\tau}\right)^{1/4}\,,
\end{align}
\end{subequations}
where $\tau=L^2/D$ is the diffusion time.

\subsection{Multiplicative growth: $L_n=\alpha\, L_{n-1}, \alpha>1$.}

The approach given above can be applied to multiplicative interface recession.
That is, upon the first resetting, the initial boundary at $x=L$ moves to
$x=\alpha L$.  In the next resetting, the boundary moves from $x=\alpha L$ to
$x=\alpha^2 L$, etc.  For this recession rule, the probability for the
$n^{\text th}$ reset event to occur at time at $t$, $R_n(t)$, is
\begin{subequations}
\begin{align}
  \label{eq:Rnte}
  R_{n}(t) &= \int_0^t dt'\, R_{n-1}(t-t')\, F(\alpha^{n-1}L , t')\,,
\end{align}
which, in the Laplace domain, becomes
\begin{align}
  \label{eq:Rnse}
  \widetilde{R}_{n}(s) &=  \widetilde{R}_{n-1}(s)   \, e^{-\sqrt{s/D}\,\alpha^{n-1}L }
     \equiv \widetilde{R}_{n-1}(s)   e^{-\ell\, \alpha^{n-1} }   = e^{-\ell\, (1-\alpha^n)/(1-\alpha)} \,.
\end{align}
\end{subequations}
The average number of reset events up to time $t$ is
\begin{subequations}
\begin{align}
\label{eq:Nte}
  \mathcal{N}(t)= 0\times Q(L,T) +  \sum_{n=1}^\infty n\int_0^t dt'\,
  Q(\alpha^n L,t-t')\, R_{n}(t') \,,
\end{align}
which becomes, in the Laplace domain, 
\begin{align}
  \label{eq:Nse}
  \widetilde{\mathcal{N}}(s)=
  \sum_{n=0}^\infty n \, \widetilde{Q}(\alpha^n L,s)\, \widetilde{R}_{n}(s) \,.
\end{align}
\end{subequations}
Substituting in $\widetilde{Q}(\alpha^n L,s)= [1-\widetilde{F}(\alpha^nL,s)]/s$ and
\eqref{eq:Rnse} into the above equation, we obtain
\begin{align}
  \widetilde{\mathcal{N}}(s)=&
    \sum_{n=0}^\infty \frac{n}{s}\, \left(1-e^{-\ell\alpha^n}\right) e^{-\ell\, (1-\alpha^n)/(1-\alpha)}\nonumber \\
\approx& \int_0^\infty dn\, \frac{n}{s} \,\left(1-e^{-\ell\alpha^n}\right) e^{-\ell\, (1-\alpha^n)/(1-\alpha)}\nonumber\\
 =&\frac{1}{s} \,\,e^{-\ell/(1-\alpha)}\int_0^\infty dn\, n\,
     (1-e^{-\ell\, \alpha^n})\, e^{\ell\, \alpha^n/(1-\alpha)}\,.
\end{align}

To evaluate the above integral, we make the variable change
$z=\ell\, \alpha^{n-1}$, from which $n= \ln(z\,\alpha /\ell) /\ln\alpha$ and
$dn=dz/(z\ln\alpha)$.  The above integral now becomes: 
\begin{align*}
  \widetilde{\mathcal{N}}(s)=&\, \frac{1}{s} \,\,\frac{e^{-\ell/(1-\alpha)}}{(\ln\alpha)^2}
    \int_{\ell/\alpha}^\infty\,\, dz \,\,\frac{\ln( z\, \alpha /\ell)}{z}\,
    (1-e^{-\alpha\, z}) \,e^{\alpha\, z/(1-\alpha)}\,, 
\end{align*}
For $s\to 0$, we compute the above integral by first splitting it into two
terms, by using $\ln( z\alpha /\ell) = \ln z + \ln( \alpha /\ell)$.  The
contribution from the first term is while the second one diverges.  Dropping
the finite term, we obtain
\begin{align}
  \widetilde{\mathcal{N}}(s) \simeq\, &\frac{\ln( \alpha /\ell)}{s}
      \,\,\frac{e^{-\ell/(1-\alpha)}}{(\ln\alpha)^2}
    \int_{\ell/\alpha}^\infty\,\, dz \,\frac{1-e^{-\alpha z}}{z}\,
     e^{\alpha z/(1-\alpha)}\,\simeq -\frac{\ln( s\tau)}{2\, s\ln\alpha} ,\quad
              s\rightarrow 0\,.
\end{align}
Note that we introduce the factor $\tau$ inside the logarithm, so that this
term is manifestly dimensionless.  Thus in the long-time limit, the average
number of resetting events scales as
\begin{align}
  \mathcal{N}(t)\simeq  \frac{\ln (t/\tau)}{2 \ln \alpha}\,.
\end{align}

\section{Expanding interval: Saddle point approximation}
\label{ap:saddlepoint}
We want to compute the following integral in the small $\ell$ limit:
\begin{align}
I(\ell)=\int_{0}^\infty du\, \exp[- \frac{1}{\ell}\, f(u) ]\,,
\label{eq:Isaddlepoint}
\end{align}
where
$f(u)=-\frac{1}{2} u^2-\frac{1}{2} \text{Li}_2\left(-e^{2 u}\right)-u \log
(2)-\frac{\pi ^2}{24}$.  Because the negative exponential is rapidly
decreasing as $\ell\rightarrow 0$, the main contribution will occur when
$f(u)$ is minimum. This function has a global minimum located at $u=0$ and
can be locally approximated by $f(u)\simeq u^3/6$.  Substituting this into
Eq.~\eqref{eq:Isaddlepoint}, we recover Eq.~\eqref{eq:saddlepoint}.

\bibliographystyle{iopart-num}

\providecommand{\newblock}{}


\end{document}